\documentclass[final]{jfm}

\usepackage{graphicx}
\usepackage{newtxtext}
\usepackage{newtxmath}
\usepackage{mathtools}
\usepackage{xcolor}
\usepackage{natbib}
\usepackage{hyperref}
\hypersetup{
    colorlinks = true,
    urlcolor   = blue,
    citecolor  = black,
}

\usepackage{multirow}
\usepackage{accents}
\usepackage{comment}
\usepackage{subcaption}
\usepackage{xcolor} 
\usepackage{epstopdf, epsfig}
\graphicspath{{figures/}}
\usepackage{cleveref} 
\usepackage{tikz}

\newcommand{\RomanNumeralCaps}[1]
\linenumbers
\graphicspath{{figures/}}

\shorttitle{SID: two-layer hydraulics and instabilities}
\shortauthor{A. Atoufi, L. Zhu, and others}

\title{Stratified inclined duct: two-layer hydraulics and instabilities}

\author{Amir Atoufi\aff{1}, Lu Zhu\aff{1} 
	\corresp{\email{lz447@cam.ac.uk}}, Adrien Lefauve\aff{1}, John R. Taylor\aff{1}, Rich R. Kerswell\aff{1}, Stuart B. Dalziel\aff{1}, Gregory. A. Lawrence\aff{2},   \and P. F. Linden\aff{1}}

\affiliation{\aff{1}Department of Applied Mathematics and Theoretical Physics, Centre for Mathematical Sciences, Wilberforce Road, Cambridge CB3 0WA, UK
	\aff{2} Department of Civil Engineering, University of British Columbia, Vancouver, BC V6T 1Z4, Canada}

\begin{document}
	
	\maketitle

	\begin{abstract}
The stratified inclined duct (SID) sustains an exchange flow in a long, gently sloping duct as a model for continuously-forced density-stratified flows such as those found in estuaries.
Experiments have shown that the emergence of interfacial waves and their transition to turbulence as the tilt angle is increased appears linked to a threshold in the exchange flow rate given by inviscid two-layer hydraulics.
We uncover these hydraulic mechanisms with (i) recent direct numerical simulations (DNS) providing full flow data in the key flow regimes (Zhu \& Atoufi \textit{et al.}, {arXiv:2301.09773}, 2023), (ii) averaging these DNS into two layers, (iii) an inviscid two-layer shallow water and instability theory to diagnose interfacial wave behaviour and provide physical insight.
The laminar flow is subcritical and stable throughout the duct and hydraulically controlled at the ends of the duct. As the tilt is increased, the flow becomes everywhere supercritical and unstable to long waves. An internal undular jump featuring stationary waves first appears near the centre of the duct, then leads to larger-amplitude travelling waves, and to stronger jumps, wave breaking and intermittent turbulence at the largest tilt angle.
Long waves described by the (nonlinear) shallow water equation are locally interpreted as linear waves on a two-layer parallel base flow described by the Taylor-Goldstein equation. This link helps us interpret long-wave instability and contrast it to short-wave (e.g. Kelvin-Helmholtz) instability.
Our results suggest a transition to turbulence in SID through long-wave instability relying on vertical confinement by the top and bottom walls.
	\end{abstract}

	\section{Introduction}\label{sec:intro} 
	
	Buoyancy-driven exchange flows between water masses of different density are common in estuaries and straits, e.g. the straits of the Great Belt, Gibraltar, Bab el Mandab, and the Bosphorous \citep{Gregg2002}. These essentially hydrostatic and two-layer flows are known to exhibit hydraulic jumps \citep{FARMER19881,thorpe_malarkey_voet_alford_girton_carter_2018}, which are important discontinuities in the internal flow properties (layer thickness and speed) and often lead to instability  \citep{lawrence_armi_2022}. A local jump can influence the large-scale properties of the flow such as the exchange flow rate \citep{lawrence_1993}. These non-local hydraulic flow features are often studied in an idealised `shallow-water' setting consisting of fluid organised in two counter-flowing, frictionless layers of specified thickness and speed with constant densities \citep{armi1986hydraulics, lawrence1990hydraulics,dalziel_1991}.  
	
	In this paper, we employ two-layer hydraulics as a diagnostic tool to derive insights from direct numerical simulation (DNS) data of the exchange flows in the stratified inclined duct (SID). SID is a canonical stratified shear flow through a long,  square cross-section, tilted duct for which there is now ample data, both experimental \citep[e.g.][]{partridge_versatile_2019} and numerical  \citep{ZhuAtoufi2022}. The investigation of turbulence in two-layer shear flows through tubes dates back to the classic experiments of \cite{reynolds_experimental_1883} and \cite{thorpe_method_1968}, who both used a closed tube. The opening of the tube into large reservoirs in SID is more recent and allows for interfacial waves and turbulence to be sustained for much longer time periods, and for various flow regimes to be distinguished. The successive transitions to increasingly turbulent regimes in SID, as the Reynolds number and tilt angle are increased, have been recognised since \cite{macagno_interfacial_1961} and \cite{kiel_buoyancy_1991}, and have been much studied more recently \citep{meyer2014stratified,lefauve2019regime,lefauve2020buoyancy,matute}. These transitions are underpinned by many fundamental features of stratified flows, including interfacial waves and turbulent intermittency.
	
	Our first aim with this new analysis is to uncover internal hydraulic effects in order to explain some of these leading-order dynamics that DNS and experimental data have revealed but not yet explained. In particular, we will provide the first direct evidence for the existence of internal hydraulic jumps, where the layer thickness expands in the direction of the flow of each layer. We also show how the development of jumps and large-amplitude interfacial waves coincides with a plateau in the exchange flow rate through the duct, after an initial increase with increasing Reynolds number and tilt in the laminar regime. This upper bound is a remarkably robust feature of SID, which has deep ramifications for the flow energetics and transition to turbulence \citep{lefauve2019regime}. Though the emergence of waves and turbulence has long been linked to the notion of `hydraulic control' in the experimental SID literature,  this link has not yet been studied in detail, primarily because experiments lack the full velocity, density and pressure data along the  length of the duct. The recent availability of DNS data \citep{ZhuAtoufi2022} overcoming these limitations finally makes a rigorous hydraulics analysis of SID possible.
	
	Our second aim is to link two-layer internal hydraulic effects to the growth of instabilities and to shorter (non-hydrostatic) waves, links which are rarely found explicitly in the literature. The streamwise variation of the base flow in the SID geometry, and generally in all exchange flows, distinguishes them from idealised parallel stratified shear flows. This variation, however, is essential to the formation of internal hydraulic jumps, to the ideas of hydraulic control and maximal exchange, and hence to the nature of SID turbulence. It is well known that under some flow conditions, the loss of hyperbolicity of the shallow-water equations renders long waves unstable. However much less is known about the consequences of this instability, and the relative importance of long-wave versus short-wave instability. We will clarify this by explaining how a certain range of unstable nonlinear shallow water waves associated with an internal jump can be interpreted as linear instabilities on a locally parallel base flow, and thus that the insights derived from stable hydraulic theory remain valid even under (moderately) unstable conditions.  
 
	To tackle these aims, we introduce our DNS datasets in \S\ref{sec:DNS}, and the two-layer averaging in \S\ref{sec:theory_2lay}, using the averaged datasets to show evidence of jumps and maximal exchange.  In \S\ref{sec:char-instab}, we adapt the two-layer shallow water theory to SID, summarise important results from the literature, and connect them to linear stability theory. In \S\ref{sec:two-lay_hydr}, we use this hydraulics and stability framework to analyse our DNS. Then, in \S\ref{sec:twoLay_LSA_TG}, we explore the transition between long (hydrostatic) waves and  short (non-hydrostatic)  waves, the influence of molecular diffusion (Prandtl number) and of smoother flow profiles. Finally, we draw our conclusions in \S\ref{sec:conclusion}.	
	
\section{Direct numerical simulations}\label{sec:DNS}

	\begin{figure}
		\centering		
		\includegraphics[width=.6\linewidth, trim=0mm 0mm 0mm 0mm, clip]{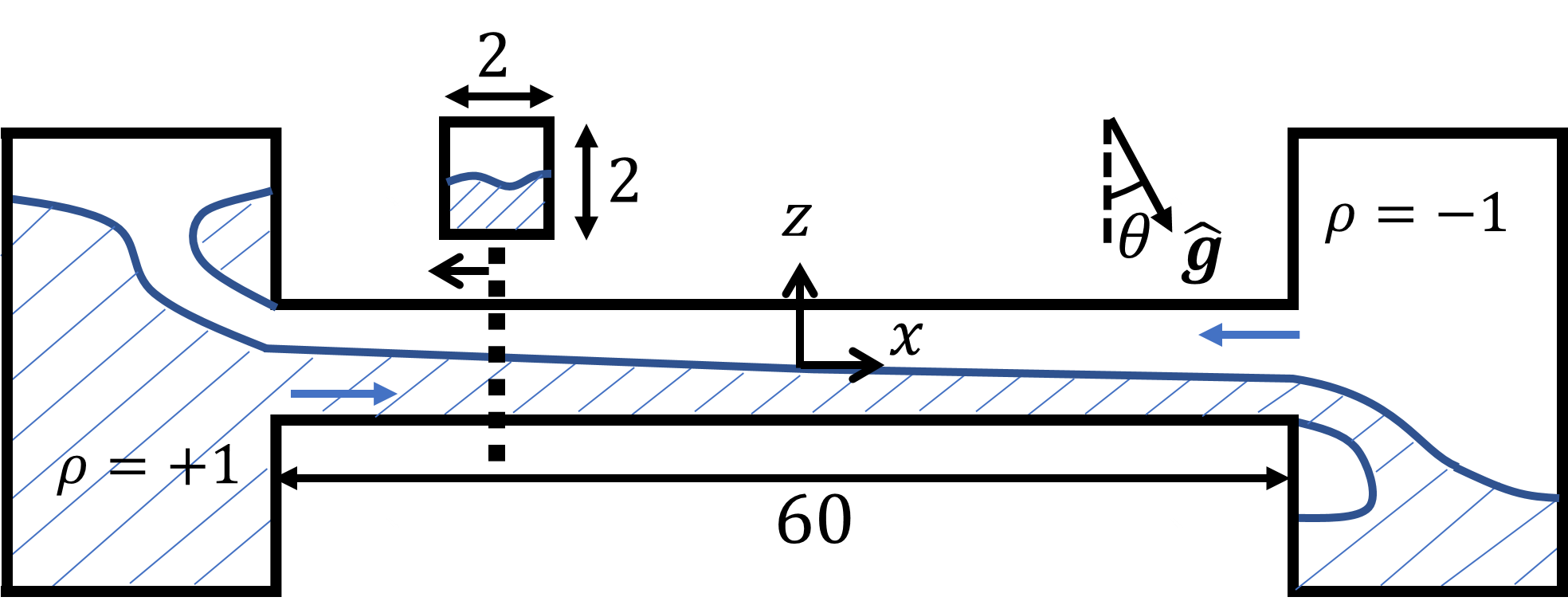}
		\caption{A schematic of the stratified inclined duct (SID) numerical  setup. The duct is oriented at angle $\theta$ to the horizontal, which is equivalent to tilting the gravity vector. Densities are non-dimensionalised by the density differences between the reservoirs  and lengths by the half-duct height. The duct's volume is $(x,y,z)\in [-30,30]\times[-1, 1]\times[-1,1]$.}\label{fig:goemetry}
	\end{figure}

\subsection{Methodology}

Our DNS solves the following non-dimensional Navier-Stokes equations under the Boussinesq approximation for the density-stratified flow in the SID setup sketched in figure~\ref{fig:goemetry}:
\begin{eqnarray}
\label{eq:ns:mass}%
\boldsymbol{\nabla} \cdot \boldsymbol{u} &=& 0,%
\\
\label{eq:ns:mom}%
\frac{\partial \boldsymbol{u}}{\partial t} + \left(\boldsymbol{u} \cdot \nabla \right) \boldsymbol{u} &=&
- \boldsymbol{\nabla}p + \frac{1}{\Rey} \nabla^{2}\boldsymbol{u} 
+ {\bf \hat{g}} \ \Ri \ \rho - \boldsymbol{F}_u,%
\\
\label{eq:ns:den}%
\frac{\partial \rho}{\partial t} + \left(\boldsymbol{u} \cdot \nabla \right) \rho &=&
\frac{1}{\Rey \ \Pran} \nabla^{2}\rho  - F_\rho,%
\end{eqnarray}
where $\Rey = H^* U^\ast/\nu$ is the input Reynolds number ($U^*=\sqrt{g'H^*}$  is the characteristic buoyancy velocity, $H^*$ is the dimensional half-height of the duct, $\nu$ is the kinematic viscosity, $g'$ is the reduced gravity); $\Ri =  g^\prime H^\ast/(2U^{\ast})^2$ is the input bulk Richardson number, giving a fixed $\Ri=1/4$; and  $\Pran\equiv \nu/\kappa$ is the Prandtl number ($\kappa$ is the thermal diffusivity). The unit gravity vector in the coordinate system $(x,y,z)$ aligned with the duct is ${\bf\hat{g}}\equiv[\sin\theta, 0, -\cos \theta]$. The DNS is performed with the open-source solver Xcompact3D~\citep{Bartholomew2020xcompact3d}.

The duct has a square cross-section of non-dimensional height and width $2$ and of length $60$ (giving a long aspect ratio of $A=30$). No-slip boundary conditions for $\boldsymbol{u}$ and no-flux boundary conditions for $\rho$ are applied on the duct walls in the spanwise ($y=\pm 1$) and vertical ($z=\pm 1$) directions with an immersed boundary method. The flow is driven by the density difference between the dense ($\rho=1$) and light ($\rho=-1$) fluid in the left- and right-hand reservoirs, respectively, producing counter-flowing layers in the streamwise $x$-direction.  The experimental reservoirs are modelled by ad hoc forcing terms $\boldsymbol{F}_u$ and $F_\rho$, which damp flow  in the reservoirs and restore the density field to $\rho=\pm 1$, allowing us to maintain a quasi-steady exchange flow for arbitrarily long times at  a minimal computational cost. Details about the numerical setup and validation against experiments and benchmark cases (with large reservoirs and $\boldsymbol{F}_u=F_\rho=0$) can be found in \citet{ZhuAtoufi2022}.

The DNS is started at $t=0$ from lock-exchange initial conditions, after which two counter-flowing gravity currents develop from $x=0$, advance at absolute speed $\approx 1$, and reach either end of the duct after $\approx 30$ advective time unit (ATU).  Shortly after, the statistically-steady exchange flow of interest in SID becomes established. Conservatively, we only retain $t\ge 80$ in the following analysis to discard any initial transients, and we run the simulation until $t=260$.

When the setup is tilted by an angle $\theta > 0^\circ$, the streamwise component of gravity contributes $\Ri \, \rho \sin\theta$ to the $x$-component of the momentum and adds extra kinetic energy to the flow. Increasing $\theta$ and/or $Re$ leads to a variety of flow regimes from laminar to wavy to intermittently turbulent to fully turbulent, found both in DNS~\citep{ZhuAtoufi2022} and experiments \citep{meyer2014stratified,lefauve2019regime,lefauve2020buoyancy}.

\subsection{Database}

We use the DNS database recently acquired by \citet{ZhuAtoufi2022}, which shows good agreement with experiments when all five non-dimensional parameters $\Rey,\Ri,\Pran,\theta, A$ are matched. This database provides the complete set of flow variables all along the duct, which is a requirement to properly study hydraulic processes in SID.

We consider flows at a fixed Reynolds number $\Rey = 650$ and fixed Prandtl number $\Pran = 7$, corresponding to temperature-stratified water. Four cases are examined, corresponding to tilt angles $\theta=2^\circ, 5^\circ, 6^\circ, 8^\circ$, denoted  by B2, B5, B6, B8, respectively in \citet{ZhuAtoufi2022}. Each of these four cases covers specific flow regimes: B2 is  laminar (L), B5 has stationary waves (SW), B6 has travelling waves (TW), and B8 has intermittent turbulence (I). In the following, we refer to these datasets as L, SW, TW, and I respectively. We only briefly touch upon more  turbulent cases (e.g. B10 having $\Rey=1000$  and $\theta=10^\circ$, referred to here as T), as they deviate significantly from the assumptions of the model in this paper, that the flow is primarily composed of two layers with a hydrostatic pressure field (discussed in \S\ref{sec:dns_pre}). For more details about the flow regimes, statistics and spatio-temporal dynamics, see \citet{ZhuAtoufi2022}.

\section{Two-layer model as a diagnostic tool}\label{sec:theory_2lay}

  \subsection{Layer averaging procedure}\label{sec:lay_ave}
  
  	\begin{figure}
		\centering		
		\includegraphics[width=.55\linewidth, trim=0mm 0mm 0mm 0mm, clip]{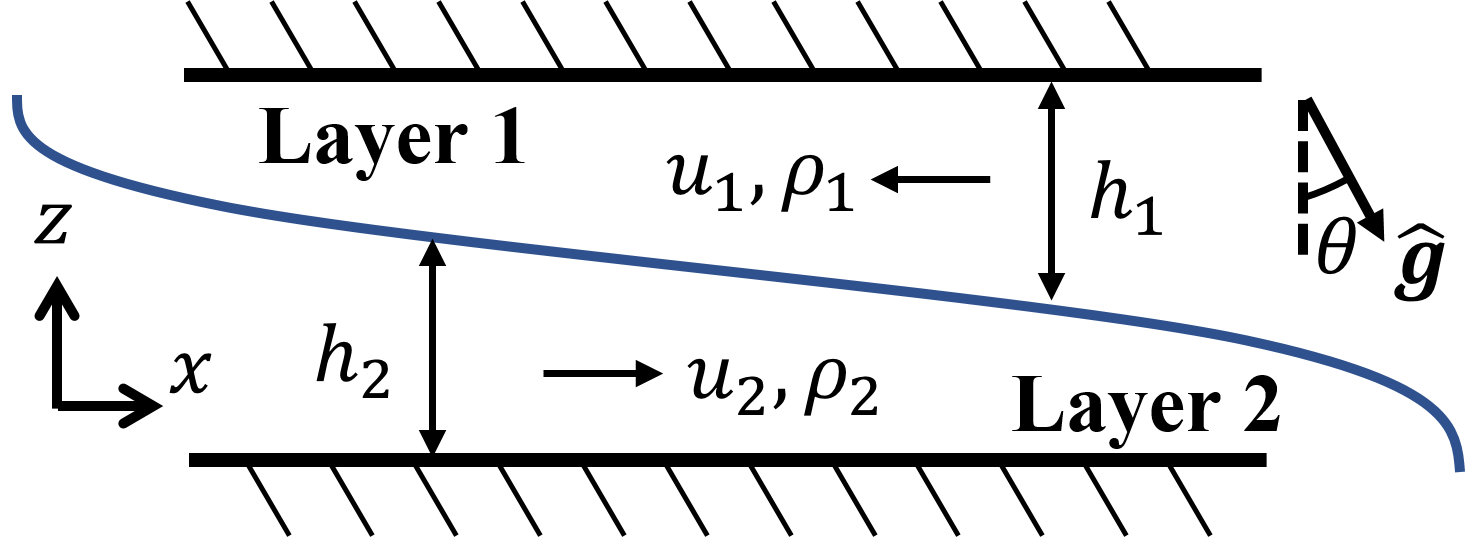}
		\caption{A schematic of two-layer shallow water flow. Flow in a part of the reservoir is considered.   }\label{fig:lay_schem}
	\end{figure}

    We seek to reduce the dimensionality of our DNS datasets to a set of two layers in order to interpret their dynamics using simple two-layer hydraulics, as sketched in figure~\ref{fig:lay_schem}. To do this, we will define the interface that separates the layers as the height $z=\eta(x,t)$ where $\rho=0$. The streamwise velocities, heights and densities of each layer are $u_i$, $h_i$, and $\rho_i$ (where $i=1,2$ correspond to the upper and lower layer, respectively), are then obtained by averaging the DNS data over the $y$-direction and in $z$ over the height of each layer. Specifically, the flow properties of the layers are obtained by
    	\begin{eqnarray}
	     h_1(x,t)=1-\eta(x,t), \quad u_1(x,t)=\langle\langle u\rangle_y\rangle_{z1},  \quad   \rho_1(x,t)=\langle\langle \rho \rangle_y\rangle_{z1}, \  
	     \\
	     h_2(x,t)=1+\eta(x,t), \quad  u_2(x,t)=\langle\langle u\rangle_y\rangle_{z2},  \quad \rho_2(x,t)=\langle\langle \rho \rangle_y\rangle_{z2}, \    
	    \label{eq:hydro_ave}
	\end{eqnarray}
	where the top-layer average is $\langle \cdot \rangle_{z1} =(1/h_1)\int_{\eta}^{1}\cdot \ dz$, the bottom-layer average is $ \langle \cdot \rangle_{z2} = (1/h_2)\int_{-1}^{\eta}\cdot \ dz$ and  and the spanwise average is   $\langle \cdot \rangle_y = (1/2)\int_{-1}^{1} \cdot \ dy$. Recall that $z=1$ and $z=-1$ are the non-dimensional height of the top and bottom walls, respectively.
    \Cref{fig:ins_flow} shows a single snapshot at time $t=110$ of $u$ (colours) and $\rho$ (contours) at the $y=0$ midplane for the two datasets L and TW, highlighting the $\rho=0$ density interface with a thick green contour.

\subsection{Layer-averaged DNS data and evidence of jumps}\label{sec:quantities}

\Cref{fig:hu_TW} illustrates the results obtained after applying this layer averaging to the DNS database. We show $x - t$ diagrams of the lower-layer height $h_2$ (top row) and lower-layer velocity $u_2>0$ (bottom row) for the laminar (L), stationary wave (SW), travelling wave (TW), and intermittently turbulent (I) cases. The layer averaging (in $y,z$) was performed in the range $|x|\leq 32$, which includes the duct $|x|\le 30$  and extends slightly into the reservoirs where the layers flow in and out. 

    \begin{figure}
    	\centering	
    	\includegraphics[width=\linewidth, trim=5mm 0mm 0mm 0mm, clip]{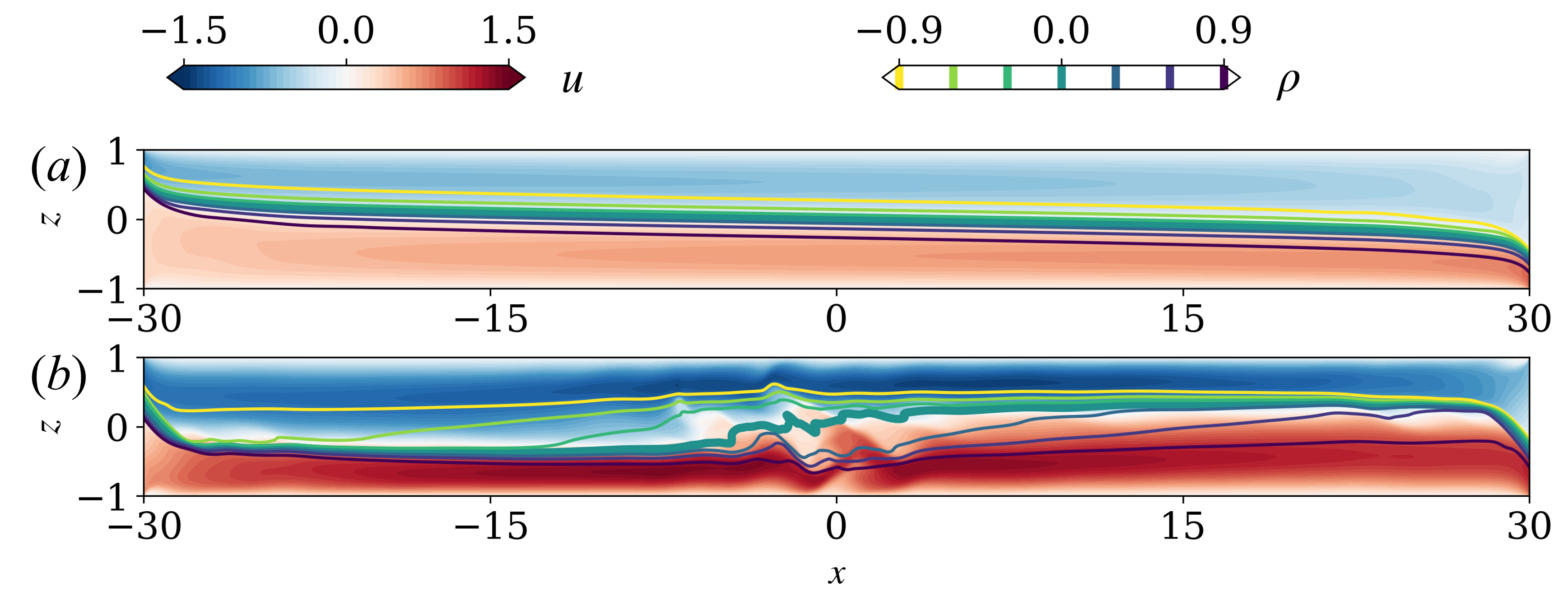}
        \caption{Snapshot at $t=110$ of streamwise velocity $u$ (blue to red shading) and density $\rho$ (colour contours) in the midplane ($y=0$) for the ($a$) laminar (L)  flow  and ($b$) travelling wave (TW) flow, the latter showing evidence of an  jump (see the interface $\rho=0$ emphasised by the thick green line).  }
        \label{fig:ins_flow}
    \end{figure}

The L flow regime exhibits sudden changes in depth and speed only at the ends of the duct $x=\pm 30$, as is expected from the flow exiting into  the  deep reservoirs. Figure~\ref{fig:hu_TW}(a) and an instantaneous snapshot of the flow in figure~\ref{fig:ins_flow}(a) confirms that the interface $\eta(x)$ is steady and gently sloping down ($\eta'<0$) throughout the duct, and is symmetric in the duct centre, i.e. $\eta(x,t)=\eta(x)=-\eta(-x)$.

In contrast, for the SW and TW flow regimes, the upper layer thickness $h_2$ suddenly increases along the direction of the flow (purple for $x<0$ to orange for $x>0$ in figure~\ref{fig:hu_TW}), which is accompanied by a sudden drop in $u_2$ (dark to light red). This discontinuity indicates the presence of what is commonly called an `internal hydraulic jump'~\citep{baines2016internal,thorpe2018application}, as can be seen in figure~\ref{fig:ins_flow}(b), where both layers experience a sudden expansion and deceleration in their respective flow direction. In the TW flow (figure~\ref{fig:ins_flow}(b))  the interface is sloping up ($\eta'>0$) in the vicinity of the jump, and is negative ($\eta<0$) throughout most of the left-hand side of the duct, and vice versa, which is the opposite of what is found in the L flow (figure~\ref{fig:ins_flow}(a)).

    \begin{figure}
    	\centering		
    	\includegraphics[width=1.03\linewidth, trim=0mm 0mm 0mm 0mm, clip]{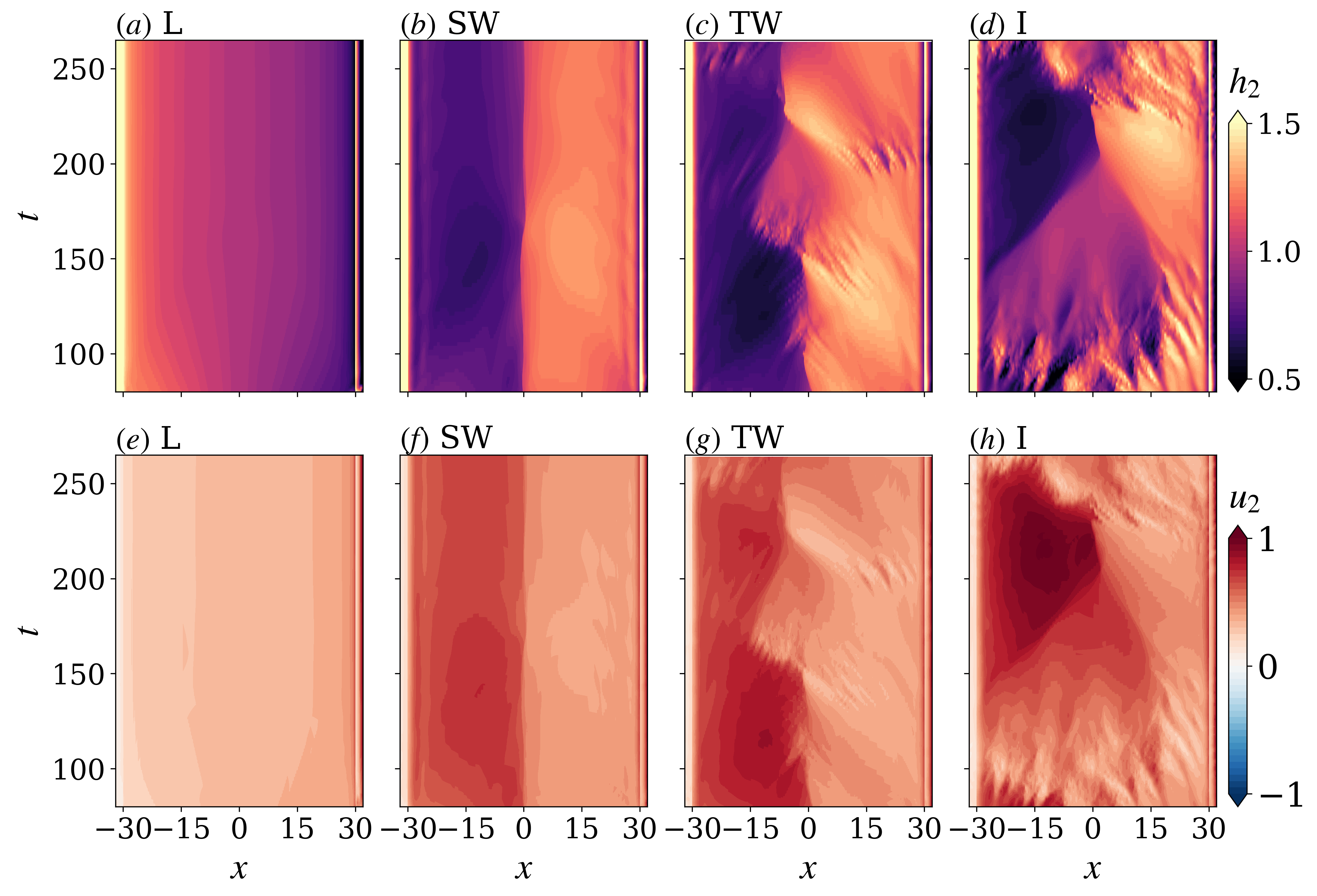}
        \caption{Spatio-temporal diagrams of (a-d) the lower-layer height $h_2$ and (e-h) lower-layer velocity $u_2$ from the DNS data for the four flow regimes (a,e) L, (b,f) SW,  (c,g) TW, (d,h) I. All data are for $t\in[80,260]$.}
        \label{fig:hu_TW}
    \end{figure}
            \begin{figure}
    	\centering	
    	\includegraphics[width=1.0\linewidth, trim=0mm 0mm 0mm 0mm, clip]{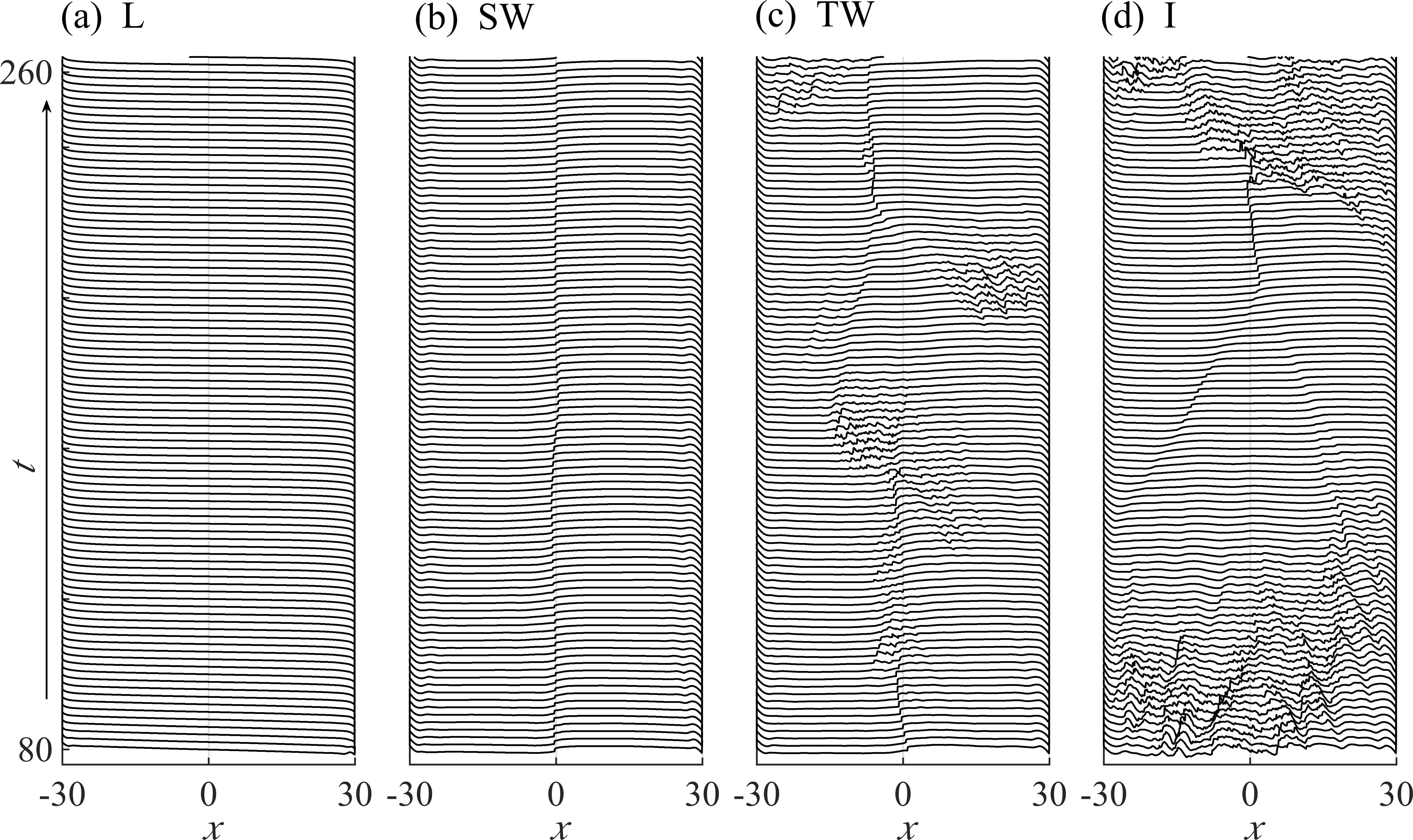}
        \caption{Temporal evolution of the density interface $\eta(x,t)$ in (a) L, (b) SW, (c) TW, (d) I. The interface curves are stacked in time at intervals of one ATU. Jumps are revealed in (b)-(d) by the discontinuity in $\eta(x)$.
        }
        \label{fig:interface}
    \end{figure}

\Cref{fig:interface} shows the temporal evolution of the interface in all four flows, with $\eta(x,t)$ plotted at intervals separated by one ATU and vertically stacked. In the SW case, the jump remains in the narrow interval $x=\pm 1$ whereas in the TW case, it oscillates over a much larger interval and sends off waves in either direction, hence the distinction between the `stationary' and `travelling' wave regime. In the I case, moving jumps are observed  in the quiet phase ($150 \lesssim t \lesssim 200$), being initiated near both ends of the duct at $t\approx150$ and progressively moving toward the centre of the duct. These jumps merge at around $t=200$ and then stay at the middle of the duct $x=0$ for $\approx20$ ATU before the transition to turbulence occurs at $t \approx 220$ (the active stage of the intermittent cycle).

\subsection{Flow rate and evidence of maximal exchange}

 To a good approximation, the flow in SID has zero net (barotropic) volume flow rate 
    \begin{equation}
   \langle u \rangle_{y,z} \equiv \frac{1}{4} \int_{-1}^{1} \int_{-1}^{1} u \, dy \, dz = \frac{1}{2}(u_1 h_1 + u_2 h_2) \approx 0,
    \label{eq:zero-net}
    \end{equation}
but a non-zero exchange volume flow rate (or volume flux) $Q(x,t)$ and mass flow rate (or mass flux) $Q_m(x,t)$ defined as 
\begin{eqnarray} 
    Q &\equiv& \langle |u| \rangle_{y,z} = \frac{1}{2}(u_2h_2 - u_1h_1) \approx -u_1 h_1 \approx u_2 h_2 \quad \text{using} \ \ \eqref{eq:zero-net}, \label{def-Q} \\
    Q_m &\equiv& \langle \rho u \rangle_{y,z}   \approx \rho_1 u_1 h_1 \approx \rho_2 u_2 h_2.\label{def-Qm}
\end{eqnarray}
The approximation in \eqref{def-Qm} comes from the fact that the layer averages of the product $\rho u$ is not exactly equal to the product $\rho_i u _i$ of the layer averages. Also, recall that the non-dimensional density is defined such that the mean density is 0 and the minimum and maximum are $-1$ and 1 respectively.

Hydraulic jumps in two-layer flows are often connected to the notion of maximal exchange, i.e. of an upper bound in the exchange volume flux $Q$ and mass flux $Q_m$. This means that flows lacking such jumps have a lower $Q$, and that no realisable flow may have a higher $Q$. While hydraulic jumps have not yet been investigated in detail in the experimental SID literature, the mass flux $Q_m$ has, due to the simplicity with which it can be measured.

	\begin{figure}
		\centering		
		\includegraphics[width=.45\linewidth, trim=0mm 0mm 0mm 0mm, clip]{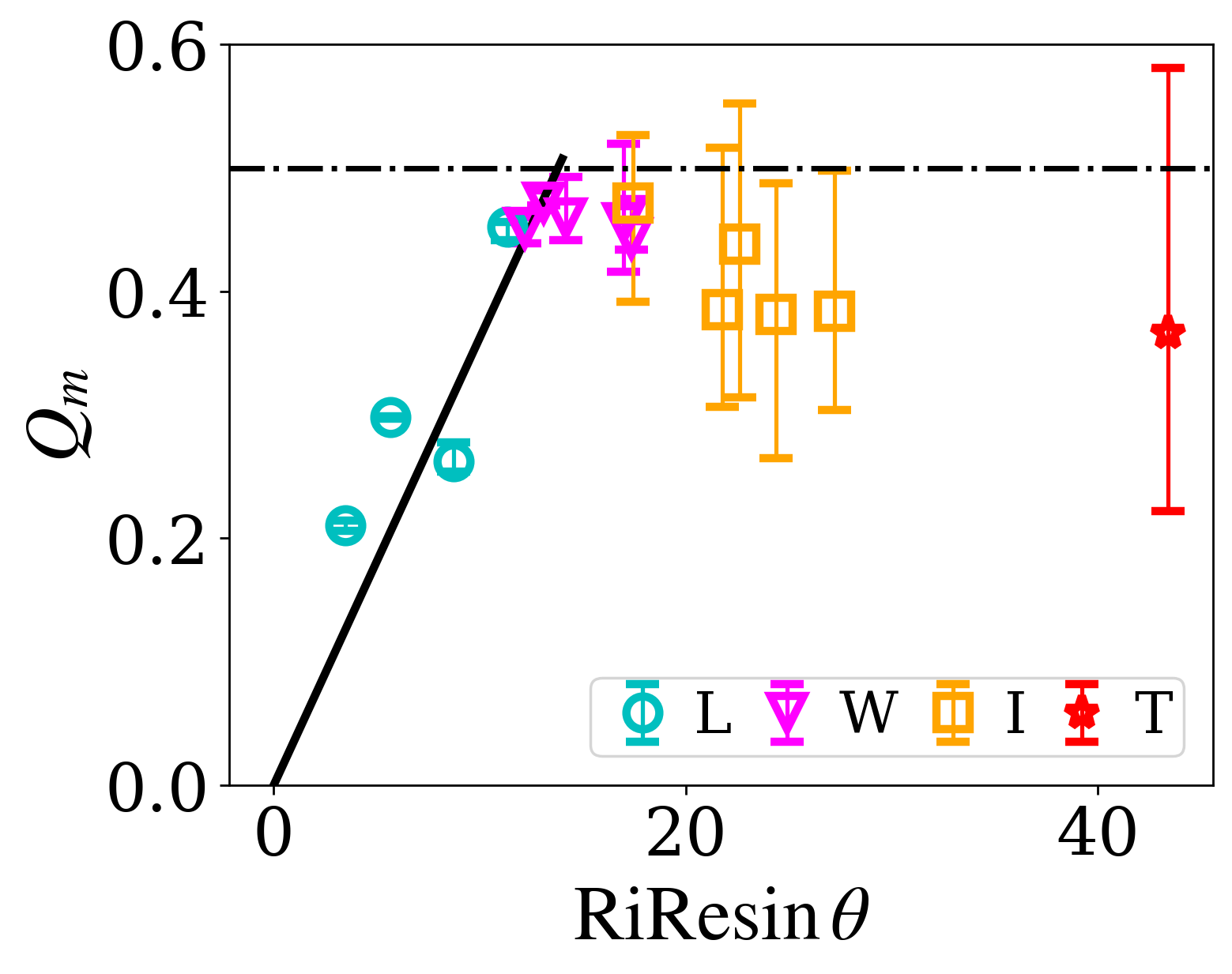}
		\caption{The mass flux $Q_m$ from the 15 DNS of \cite{ZhuAtoufi2022} increasing with $Re$ and $\theta$ until the upper bound of $\approx 0.5$. The symbols, coded by the respective regime, show the time-averaged value while the error bars depict the extreme values}\label{fig:Qm}
	\end{figure}

In figure~\ref{fig:Qm} we show the dependence of the mass flux  $Q_m$ with $\Ri \Rey \sin \theta \approx (1/4) \theta  \Rey$  using its temporal mean (symbols) and extreme values (error bars) in 15 different DNS from \cite{ZhuAtoufi2022}, containing the four main datasets L, SW, TW and I, as well as 11 others including one turbulent dataset (T). We find that $Q_m$ increases approximately linearly with $\theta \Rey$ in the L regime (where little mixing ensures that $Q\approx Q_m$), in agreement with the laminar analytical solution of \cite{matute}. For higher values of $\theta \Rey$ it reaches an upper bound $Q_m\approx 0.5$ in the W regime, before dropping below $0.5$ in the I and T regimes. These observations are consistent with the corresponding experimental data of \cite{lefauve2020buoyancy} (their figures 5-6), if we exclude their data at small angles $\theta<2^\circ$ (which behave slightly differently and we did not simulate).  This apparent upper bound is an evidence of maximal exchange, since further increase in $\theta$ does not lead to an increase in the velocity difference between the upper and lower layer. It is a significant departure from  the laminar solution based on the balance of gravitational forcing and viscous drag, suggesting $Q_m\propto \theta \Rey$ \citep{lefauve2020buoyancy}. As it will be shown later (\cref{sec:maxExchange}) this $Q_m \approx 0.5$ is the threshold of the long-wave instabilities and onset of hydraulic transitions and a further increase in $Q_m$ beyond this limit is taxed by turbulent dissipation.

This large body of experimental and numerical evidence in favour of maximal exchange in SID and $Q_m\approx 0.5$ at the transition between the L and W regime, combined with our observations in \S\ref{sec:quantities}  of the existence of an internal jump, is strongly suggestive that hydraulic effects dominate the flow in the W regime onwards. 

In the next section, we develop the two-layer hydraulics framework to study these jumps and maximal exchange and their relation to the transition from laminar flow to waves and to turbulence in SID.

\section{Two-layer equations: characteristics and instabilities} \label{sec:char-instab}

In this section, we aim to define and relate the characteristic velocity of the two-layer flows in the context of SID to the hydraulic regime of the flow. We also aim to provide a physical implication of when this characteristic velocity is not purely real. We start by simplifying the inviscid Navier-Stokes equations with shallow water (long wave) theory in \S\ref{sec:SWE}-\ref{sec:G2} before comparing it to the Taylor-Goldstein (linear wave) theory in  \S\ref{discontinous_2L}, and interpreting our findings in \S\ref{sec:link}.

\subsection{Shallow water equations: nonlinear long waves}\label{sec:SWE}
    
The validity of this hydrostatic assumption is verified in our DNS data in \cref{sec:dns_pre}. The upper layer then obeys  
     \begin{eqnarray} \label{2L_SWE_L1}
    &&\frac{\p h_1}{\p t} + \frac{\p (h_1 {u}_1)}{\p x}=0,\\&& \nonumber
\frac{\p (h_1 {u}_1)}{\p t} + \frac{\p   }{\p x}\left(h_1{u}_1^2+ \Ri \cos{\theta} {\rho}_1 \frac{h_1^2}{2}\right)= \Ri \sin{\theta} {\rho}_1 h_1 -\Ri \cos{\theta} h_1 \frac{\p}{\p x} \left(p_w+{\rho}_1 h_2+
    {\rho}_1 b \right),
    \end{eqnarray}
    and  the lower layer obeys
    \begin{eqnarray}\label{2L_SWE_L2}
    && \frac{\p h_2}{\p t} + \frac{\p (h_2 {u}_2)}{\p x}=0,\\&& \nonumber
    \frac{\p (h_2 {u}_2)}{\p t} + \frac{\p   }{\p x}\left(h_2{u}_2^2+ \Ri \cos{\theta} {\rho}_2 \frac{h_2^2}{2}\right)  = \Ri \sin{\theta} {\rho}_2 h_2 -\Ri \cos{\theta} h_2 \frac{\p}{\p x} \left(p_w+{\rho}_1 h_1+ {\rho}_2 b \right).
    \end{eqnarray}
    Here, $p_w(x,t)$ is the pressure at the upper wall and $b(x)$ is the elevation of the bottom wall.    Conveniently, the bottom wall in SID is fixed at $b(x)=-1$. We subtract the momentum equations in \eqref{2L_SWE_L1} to \eqref{2L_SWE_L2} to remove $p_w$ and reduce the number of unknowns. The variation of density of the upper and lower layers in $x$ is also neglected compared to variations in height of the layers. The shallow water equations become  
	\begin{eqnarray} 
    \frac{\p h_2}{\p t} -\frac{\p h_1}{\p t} + h_2 \frac{\p u_2}{\p x}- h_1 \frac{\p u_1}{\p x}+u_2 \frac{\p h_2}{\p x} -u_1 \frac{\p h_1}{\p x}=0,
    \end{eqnarray}
    \begin{eqnarray}\label{eq:du2-du1}
    \frac{\p u_2}{\p t} -\frac{\p u_1}{\p t} + u_2 \frac{\p u_2}{\p x}-u_1 \frac{\p u_1}{\p x} +\Ri \cos{\theta} (\rho_2-\rho_1) \frac{\p h_2}{\p x} && = \Ri \sin{\theta} (\rho_2-\rho_1)  \\ && \nonumber  \quad -\Ri \cos{\theta}(\rho_2-\rho_1) \frac{\p b}{\p x},
    \end{eqnarray}
    with two auxiliary equations to satisfy the no-net (barotropic) flow condition and geometric constraint inside the duct $({\p b}/{\p x}=0)$
    \begin{eqnarray} \label{eq:auxiliary}
    u_1 h_1 + u_2 h_2 = 0, \quad h_1 + h_2 = 2. 
    \end{eqnarray}
    By taking the derivative of (\ref{eq:auxiliary}) with respect to $x$, these four equations can be written compactly as
    \begin{equation} \label{eq:SWE_main}
    	\mathsfbi{C}\frac{\p \boldsymbol{q}}{\p t}+\mathsfbi{A}(\boldsymbol{q},\Delta \rho)\frac{\p \boldsymbol{q}}{\p x}=\boldsymbol{f},
    \end{equation}
	where the state vector $\boldsymbol{q}$ and coefficient matrices $\mathsfbi{A}, \mathsfbi{C}$ are
	\begin{eqnarray}\label{eq:SWE_matrices}
		\boldsymbol{q}=\left[\begin{array}{llcc}
			u_1 \\
			u_2 \\
			h_1 \\
			h_2
		\end{array}\right], \
		\mathsfbi{A}=
		\left(\begin{array}{cccc}
			-u_1  & u_2  & 0  & \Ri\cos\theta\,\Delta\rho \\
			-h_1  & h_2  & -u_1  & u_2 \\
			0 & 0 & 1 & 1\\
			h_1  & h_2  & u_1  & u_2 
		\end{array}\right), \
		\mathsfbi{C}=
		\left(\begin{array}{cccc}
			-1  & 1  & 0  & 0 \\
			0  & 0  & -1  & 1 \\
			0 & 0 & 0 & 0\\
			0  & 0  & 0  & 0 
		\end{array}\right), \quad
	\end{eqnarray}
The shallow water equation (SWE) \eqref{eq:SWE_main} is quasilinear, i.e. linear in the derivatives of $\boldsymbol{q}$ but with the coefficient matrix $\mathsfbi{A}$ dependent on $\boldsymbol{q}$. The quasi-constant local density difference between layers is defined as $\Delta\rho(x,t) \equiv \rho_2-\rho_1\in (0,2)$ (specified by our layer-averaged DNS data).
This equation does not assume that interfacial waves have infinitesimal amplitudes, but it does assume, through the hydrostatic assumption, that waves are long with respect to the layer height (i.e. that their non-dimensional wavenumber $k\ll 1$). In the following, we neglect the forcing 
\begin{eqnarray}\label{eq:forcing-term}
   \boldsymbol{f}=[\Ri\sin\theta\, \Delta\rho,\, 0, \, 0,\, 0]^T,  
\end{eqnarray}
recalling that $\sin\theta\ll 1$. We will study \eqref{eq:SWE_main} in the homogeneous limit $(\boldsymbol{f}=\boldsymbol{0})$, with a focus on the eigenvalues of $\mathsfbi{A}$, in which $\boldsymbol{f}$ plays no role. The role of forcing in shallow water theory is an interesting question left for future work. However, the forcing proportional to $\sin\theta$ does influence the DNS, and is thus implicitly taken into account in the state vector $\boldsymbol{q}$ obtained after layer averaging the DNS data.

\subsection{Characteristic curves and propagation of information} \label{sec:twoLayer_def}

    Consider a left eigenvector $\boldsymbol{v}$ and eigenvalue $\lambda$ associated with the matrix pair $(\mathsfbi{A}$,$\mathsfbi{C})$ such that $\boldsymbol{v}^H \mathsfbi{A} = \lambda \boldsymbol{v}^H \mathsfbi{C}$, where $^H$ denotes Hermitian transpose. Multiplying (\ref{eq:SWE_main}) by $\boldsymbol{v}^H$ yields
    \begin{eqnarray} \label{SWE_main_f1}
    \boldsymbol{v}^H \mathsfbi{C} \left(\frac{\p \boldsymbol{q}}{\p t}+\lambda(x,t) \frac{\p \boldsymbol{q}}{\p x}\right)=\boldsymbol{0}.
    \end{eqnarray}
    The eigenvalues $\lambda$ are called characteristic velocities, since they define characteristics curves $s$ in the $(x,t)$ plane along which the partial differential equation (\ref{eq:SWE_main}) reduces to an ordinary differential equation \citep{whitham2011linear}. For this homogeneous equation,  the combinations of flow variables $\boldsymbol{v}^H\mathsfbi{C} \, d\boldsymbol{q}/ds = \boldsymbol{0}$ is conserved.
    These characteristic velocities,  henceforth simply referred to as `characteristics',  can be complex $\lambda=\lambda^R + i \lambda^I \in \mathbb{C}$. Their real part $\lambda^R \equiv \Re{(\lambda)}$ represents the phase speed of shallow water waves, describing the trajectories $s$. 
    Their imaginary part $\lambda^I \equiv \Im{(\lambda)}$ represents any potential growth ($\lambda^I>0$) or decay ($\lambda^I<0$) of these waves. The direction of information propagation is set by the sign of $\lambda^R$: when $\lambda^R>0$, information propagates rightward (towards increasing $x$), and vice versa, whereas  $\lambda^R=0$, indicates stationary waves. 
    
    The characteristics $\lambda$ are given by the two distinct solutions of $\det(\mathbi{A}-\lambda \mathbi{C})=0$:
	\begin{eqnarray}\label{2L_Char_exact}
		 \lambda_{1,2}(x,t)&=&\underbrace{\frac{h_1 u_2 +h_2 u_1 }{h_1 +h_2 }}_{\text{convective velocity}\ \equiv\ \bar{\lambda}} \pm  \ \ \underbrace{\sqrt{\frac{\Delta\rho  \Ri \, \cos{\theta} \, h_1 h_2}{h_1 +h_2} {\left(1-F_{\Delta}^2 \right),}}}_{\text{phase speed}\ \equiv \ \delta \lambda}  \hspace{3cm} \\
   &\approx& \ \qquad \overbrace{\frac{-2\eta Q}{1-\eta^2}} \quad  \pm \quad \ \overbrace{\sqrt{\frac{\Delta\rho}{2}\cos\theta(1-\eta^2)(1-F_\Delta^2)}} \quad \text{using} \ \eqref{def-Q}, \label{2L_Char_approx} \end{eqnarray}
where
   \begin{eqnarray}
	    F_{\Delta}^2(x,t)&=&\frac{({u}_2-{u}_1)^2}{\Delta\rho \ \Ri \  \cos{\theta} \ (h_1 +h_2)}, \label{2L_Fr} \\
     &\approx& \frac{2}{\Delta\rho \cos \theta}\Big(\frac{2Q}{1-\eta^2}\Big)^2 \quad \text{using} \ \eqref{def-Q}.  \label{2L_Fr_approx} 
   \end{eqnarray}

	Here,  \eqref{2L_Char_approx} and \eqref{2L_Fr_approx} use the volume flux $Q>0$ in \eqref{def-Q} (which is an approximation relying on \eqref{eq:zero-net}) and the interface position instead of  the layer heights and velocities, as well as the fact that $\Ri=1/4$ and $h_1+h_2=2$ in SID. We see that the characteristics  consist of a convective velocity $\bar{\lambda}$ and a phase speed $\delta\lambda$, which can be imaginary, depending on the value of the `stability Froude number' $F^2_{\Delta}$  \citep{long1956long,lawrence1990hydraulics, dalziel_1991}. Note that \eqref{eq:SWE_matrices}, \eqref{2L_Char_exact} and \eqref{2L_Fr} are slightly modified versions of those given in previous studies~\citep{long1956long, armi1986hydraulics,lawrence_1993} adapted to SID flows.

If $F_{\Delta}^2<1$ the two characteristics  $\lambda_{1,2}=\bar\lambda\pm \delta\lambda$ are real and information propagates in both directions relative to the convective velocity $\bar{\lambda}$. The absolute direction of propagation is given by the sign of $\lambda_{1,2}$, which we shall return to in \S\ref{sec:G2}. 

If $F_{\Delta}^2>1$ the characteristics become complex conjugates  $\lambda_{1,2} =\lambda^R\pm i \lambda^I = \bar{\lambda}\pm i|\delta\lambda|$, indicating that the system is no longer hyperbolic and that waves grow temporally unstable. The condition $F_{\Delta}^2>1$ is sometimes known as Long's instability criterion \citep{long1956long}, although it is quoted in \cite{lamb} and likely dates back to Helmholtz.   The real part is the convective velocity, i.e. information only propagates in one direction, and the growth rate is $\lambda^I=|\delta\lambda|=\sqrt{(\Delta\rho/2)\cos\theta (1-\eta^2)(F_\Delta^2-1)}$.

\begin{figure}
    \centering
       \includegraphics[width=.75\linewidth, trim=0mm 0mm 0mm 0mm, clip]{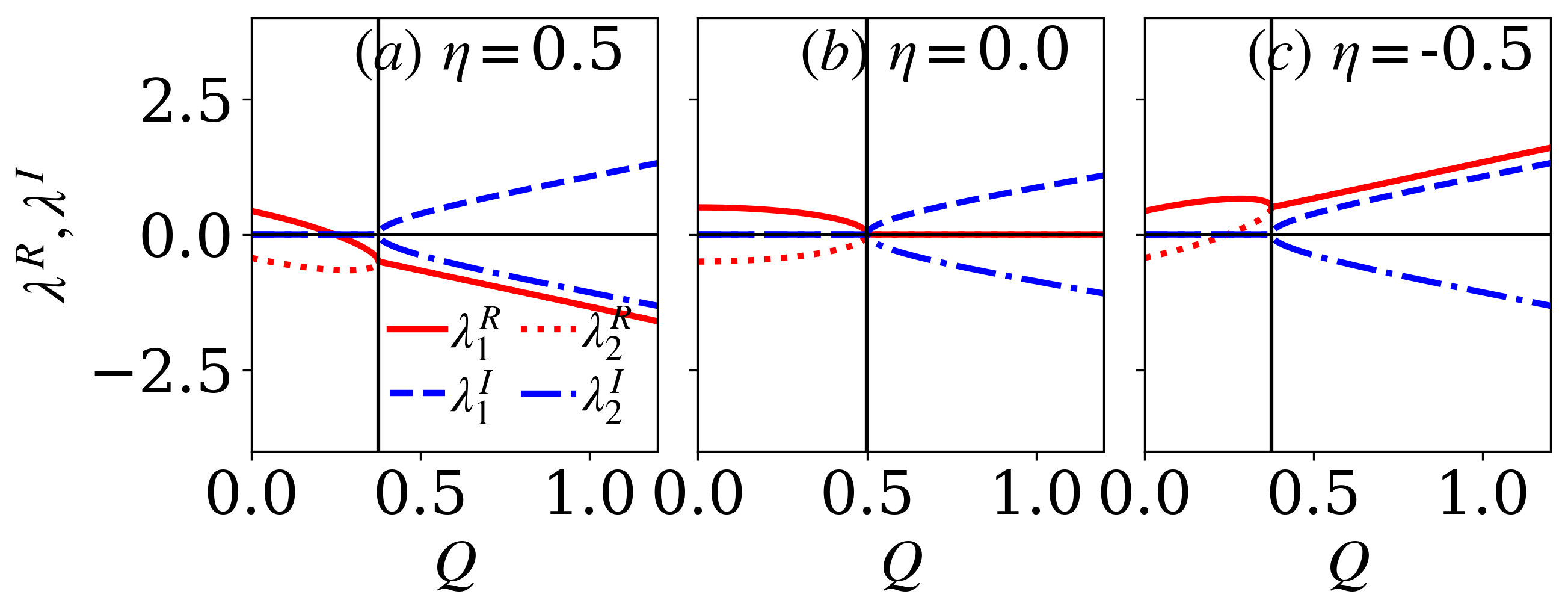}
    		\caption{Real and imaginary parts of the eigenvalues of the two-layer SWE in \eqref{2L_Char_approx} at (a) positive (b) zero and (c) negative interface elevations $\eta$ as functions of the volumetric flow rate $Q$. The eigenvalues are always complex for $Q\ge Q_c=0.5$, and become complex for slightly lower critical values $Q_c\approx 0.4$ in the asymmetric cases (a,c). }
		\label{fig:twolay_theo}
\end{figure}

 \Cref{fig:twolay_theo}(a-c) shows how $\lambda_{1,2}$ vary with the volume flux $Q$ and interface position $\eta$ following \eqref{2L_Char_approx}, assuming no mixing  ($\Delta \rho=2$) and a horizontal duct ($\cos\theta=1$). We compare the case where the interface is locally symmetric ($\eta=0$, panel b) to the cases where it is asymmetric, i.e. above  or below the mid-level ($\eta=\pm 0.5$ in panels a and c, respectively).    
Panel b shows that with a symmetric interface $\lambda_{1,2}=\pm \sqrt{1-4Q^2}$ are real (red curves) for $Q\le0.5$ and become purely imaginary (blue curves) for $Q>0.5$. However, when the interface is either below or above the midplane ($\eta=\pm 0.5$, panels a,c), this transition to instability occurs at a lower critical volume flux $Q>Q_c \approx 0.4$. In summary, instability is caused, for a given volume flux $Q$, by an increasingly asymmetric interface $|\eta|$, and vice versa, for a given interface position, by an increasing volume flux $Q$. This offers a possible explanation for the transition from L to W flow observed in figure~\ref{fig:Qm}.

\subsection{Composite Froude number and hydraulic control }
\label{sec:G2}

To further stress the importance of the characteristics $\lambda_{1,2}$, we return to the original SWE (\ref{eq:SWE_main}), and note that a non-trivial steady solution  $\boldsymbol{q}(x)$ requires $\det \mathsfbi{A} \neq 0$, i.e.
	\begin{equation} \label{eq:G_detA_com}
		\det \mathsfbi{A}=2  \Delta\rho  h_1  h_2
		\Ri  \cos{\theta}  (G^2-1) \neq 0 \quad \Leftrightarrow \quad  G^2  \neq 1
	\end{equation}
where $G^2$ is the  squared composite Froude number  defined with the Froude numbers of the upper and lower layers, $F_1$, and $F_2$, respectively,
	\begin{equation} \label{eq:G_F1_F2}
		G^2=F_1^2+F_2^2, \qquad F_i=\frac{{u}_i}{\sqrt{\Delta\rho \, \Ri \, \cos{\theta} \, h_i}} \quad i=1,2.
	\end{equation} 
Points at which $G^2=1$ are called control points \citep{armi1986hydraulics, lawrence1990hydraulics, dalziel_1991}.  At control points, $\mathsfbi{A}$ is non-invertible and a regularity condition  must exist to recover a steady solution. 

\begin{figure}
	\centering		
	\includegraphics[width=0.9\linewidth, trim=0mm 0mm 0mm 0mm, clip]{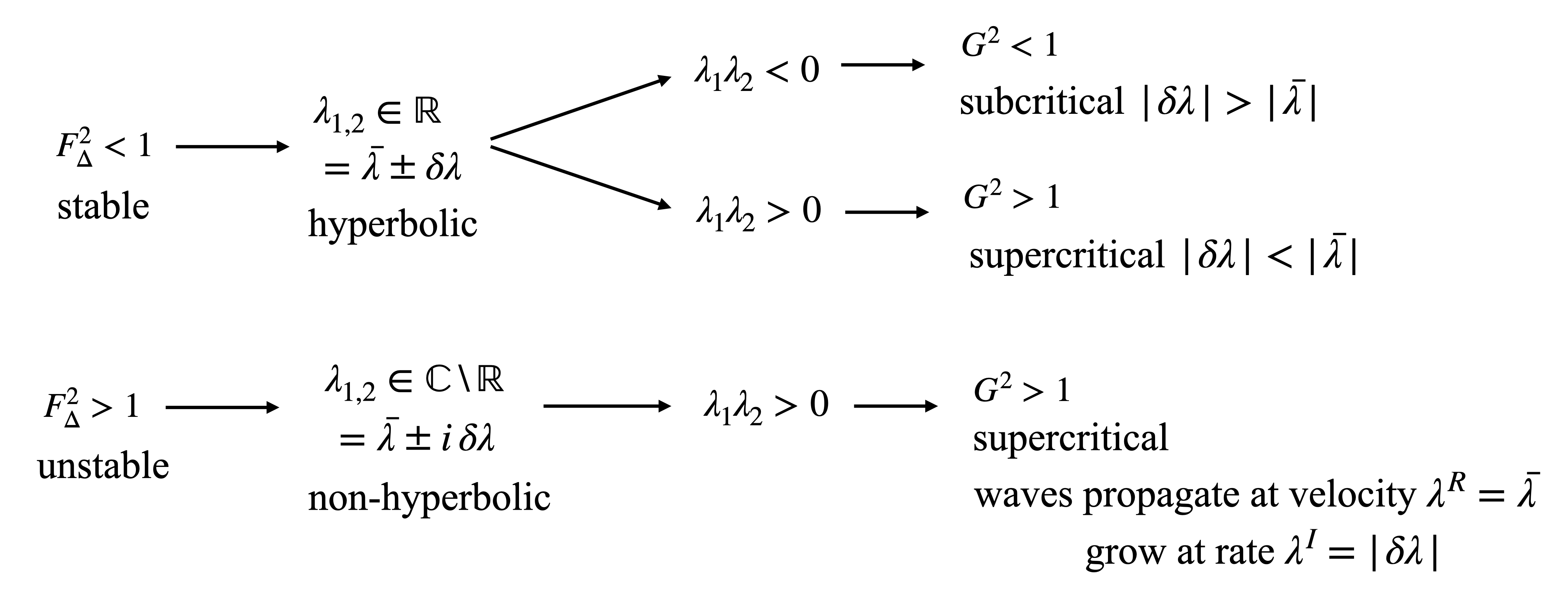}
	\caption{Summary of the scenarios revealed by the stability Froude number $F^2_\Delta$ \eqref{2L_Fr}, characteristics   $\lambda_{1,2}$ \eqref{2L_Char_exact}, and composite Froude number $G^2$ \eqref{Gcomp_Lamab2}. }
	\label{fig:link_F_G}
    \end{figure}

The link between characteristics and the composite Froude number can be highlighted with the identity
	\begin{eqnarray} \label{Gcomp_Lamab2}
		G^2=1+\frac{h_1+h_2}{({\rho}_2-{\rho}_1) \Ri \cos{\theta} \ h_1 h_2} \lambda_1 \lambda_2 &=& 1+\frac{2\lambda_1\lambda_2}{\Delta\rho\cos\theta(1-\eta^2)} \\
  &\approx & 1+\sqrt{\frac{2}{\Delta\rho \cos\theta}}\frac{\lambda_1\lambda_2F_\Delta}{2Q} \quad \text{using} \ \eqref{2L_Fr_approx} \quad
	\end{eqnarray}
From this expression, we deduce the following, illustrated in figure~\ref{fig:link_F_G}.

If the waves are stable ($F_\Delta^2<1$), the characteristics $\lambda_{1,2}$ are real. The flow is called subcritical and information propagates in both directions (along positive and negative $x$) i.e. $\lambda_1\lambda_2<0 \Leftrightarrow  G^2<1 $. In other words, the absolute phase speed $|\delta\lambda|$ is larger than the absolute convective velocity $\bar{\lambda}$ in \eqref{2L_Char_exact}. Vice versa, the flow is called supercritical  when information propagates in only one direction i.e. $\lambda_1\lambda_2>0 \Leftrightarrow G^2>1$, and the absolute phase speed $|\delta\lambda|$  is smaller than the absolute convective velocity $\bar{\lambda}$. Note that for supercritical flow, the direction of propagation associated with $\lambda_1$ and $\lambda_2$ is the same as that given by the convective velocity $\bar\lambda$, i.e. the waves are swept downstream.

If, on the other hand, the waves are unstable ($F_\Delta^2>1$), the characteristics are complex conjugates $\lambda_{1,2}=\lambda^R \pm i \lambda^I$, i.e. information always propagates in a single direction given by the sign of $\lambda^R=\bar\lambda$ and the flow is always supercritical  ($\lambda_1 \lambda_2=(\lambda^R)^2+(\lambda^I)^2>0 \Leftrightarrow G^2>1$). Under unstable conditions, control points where $G^2=1$ cannot exist, but points can exist where stationary ($\lambda^R=0$) waves grow ($\lambda^I>0$) and control the flow.

In the next sections, we clarify the interpretation of unstable shallow water waves using linear stability analysis around a locally parallel base flow. In \S\ref{discontinous_2L} we take the long-wave limit of solutions of the Taylor-Goldstein equations, and in \S\ref{sec:link} we linearise the shallow water equations assuming the waves are sufficiently (but not excessively) long.

\subsection{Taylor-Goldstein equations: linear short and long waves} \label{discontinous_2L}
    
    We relax the previous restriction that waves must be long ($k \ll 1$) by studying waves of possibly larger $k$ but of infinitesimal amplitude developing on a parallel base flow described by a velocity profile $\mathcal{U}(z)$ and a density profile $\mathcal{R}(z)$. The perturbation streamfunction $\hat{\psi}(z) \exp ik(x-ct)$ describing the evolution of these two-dimensional linear waves is given by the inviscid  Taylor-Goldstein  equation (TGE)
    \begin{eqnarray} \label{TG_eq}
    (\mathcal{U}-c) \left(\frac{{d}^2 }{{d} z^2}-k^2\right)\hat{\psi}-\mathcal{U}'' \, \hat{\psi} - \frac{\Ri \cos{\theta} \ \mathcal{R}'}{\mathcal{U}-c} \hat{\psi} = 0.
    \end{eqnarray}
    This can be analysed following standard methods \citep[e.g.][]{drazin2002introduction,smyth_carpenter_2019} with details given in \cref{sec:TG2}. Note that we assumed small tilt angles $0<\sin{\theta} \ll \cos{\theta}$, although the more general TGE in \cref{sec:TG2} 
    shows that $\sin\theta$ has a destabilising effect (ignored here). In the ordinary differential equation above, `$\prime$' denotes differentiation with respect to $z$, and $c \in \mathbb{C}$ is the phase speed of the plane waves, akin to the characteristics $\lambda$ of the SWE. However, we use a different notation for the characteristics of shallow water nonlinear waves and the phase speed of TG linear waves to emphasise that while the former implicitly assumes $k\ll 1$, the latter implicitly assumes  $k \gg A^{-1}$. In other words, the TG linear waves we investigate should be shorter than the duct length $A$ for the local analysis on a parallel base flow  to be sensible. In other words, $k \gg A^{-1}$ ensures that the waves do not `feel' the streamwise variations of the base flow along the duct.

    To make analytical progress and obtain solutions for $c$, we assume a two-layer flow bounded by solid walls, with fixed layer heights $h_1, h_2$,  velocities $u_1,u_2$, and an interface at $z=0$, consistent with the two-layer model adopted throughout this paper
	\begin{eqnarray} \label{eq:baseflow_TGE}
   \mathcal{U}(z)=
    \begin{cases}
        u_1 &  0 < z \le h_1 \\
        u_2 &  -h_2 \le z < 0
    \end{cases}, \qquad 
    \mathcal{R}(z)=
    \begin{cases}
        \rho_1 &  0 < z \le h_1 \\
        \rho_2 &  -h_2 \le z < 0
    \end{cases}.
    \end{eqnarray}
    Note that $h_1+h_2=2$ and, by a simple vertical shift, the above model is equivalent to having a domain restricted $z=\pm 1$ and an interface at an arbitrary $z=\eta$ (giving $h_{1,2}=1\pm\eta$). 
    
	By enforcing the matching conditions for the stream function and pressure at the interface (see \cref{sec:TG2}), we derive the dispersion relation for the complex phase speeds
	\begin{eqnarray} \label{lambda_2L_TG_mt}
    && c_{1,2}=\frac{u_1+u_2}{2} + \frac{ (\sigma_5-\sigma_6) (u_1-u_2) \pm \sigma_2 }{2\,{\left(\sigma_1 -1\right)}}, \\ && \qquad \nonumber
    \sigma_1 =\cosh \left(4\,k\,\right) , \ \quad \sigma_2 =\sinh \left(2k\,\right) \Lambda, \ \quad \sigma_3 =\sinh \left(2 \,k \,h_2\right), \\ && \qquad \nonumber
     \sigma_4 =\sinh \left(2 \,k \,h_1\right), \quad \sigma_5 =\cosh \left(2 \,k\,h_2\right), \quad \sigma_6 =\cosh \left(2 \,k \,h_1 \right), \\ && \qquad \nonumber 
     \Lambda =\sqrt{-\frac{4}{k} {{\left(k\sigma_4 \sigma_3 [ {u_1 }-{u_2 }]^2 + \Ri \cos{\theta} [\sigma_4(1-\sigma_5)+\sigma_3(1-\sigma_6)] \Delta\rho  \right)}}}.
     \end{eqnarray}
These waves are the well-known Kelvin-Helmholtz (KH) waves supported by a single vortex sheet (see e.g. \cite{smyth_carpenter_2019} \S~4.6.1). They are modulated by stratification; increasing  $\Ri \cos\theta \Delta  \rho $ always stabilises them.  Importantly, the dispersion relation \eqref{lambda_2L_TG_mt}  describes waves in a domain bounded by the top and bottom walls, which as we will see, strongly affect waves that have a wavelength comparable to or longer than the domain height.

For `short' waves, i.e. having a wavelength of the order of the duct height $k=O(1)$ or shorter $k>1$, the dispersion relation (\ref{lambda_2L_TG_mt}) cannot be simplified further, and these waves are dispersive. These will be referred to simply as KH waves, as in the short wave limit they are identical to those found in vertically unbounded domains.

For long waves (which, for clarity, we do not call KH waves), i.e. having a wavelength much longer than the duct height, but still shorter than the duct length $A^{-1} \ll k \ll 1$, the dispersion relation (\ref{lambda_2L_TG_mt})  simplifies as $\sinh{k h} \longrightarrow k h$ and $\cosh{k h} \longrightarrow 1$  and \eqref{lambda_2L_TG_mt} reduces to \eqref{2L_Char_exact}, i.e.
    \begin{equation} \label{eq:limit}
        c_{1,2} (k) \longrightarrow \lambda_{1,2} \quad \text{when} \ k\ll 1,
    \end{equation}
    and these waves are non-dispersive. In other words, the characteristics $\lambda_{1,2}$ of nonlinear shallow water waves can be identified to the phase speeds $c_{1,2}$ of linear (infinitesimal) long waves calculated assuming the local two-layer flow  $(u_1,u_2,h_1,h_2)$ to be parallel and steady as in \eqref{eq:baseflow_TGE}.  

The dispersion relation \eqref{lambda_2L_TG_mt} shows that there exists a smooth transition between short KH waves and long shallow water waves, as those waves differ by their wavelength but not the underlying physical mechanism. We return to this smooth transition and plot the dispersion relation in \S\ref{sec:twoLay_LSA_TG}.

Although the limit \eqref{eq:limit} can be inferred from the PhD thesis of \citet{Gu_2001} (\S~3.3) and is briefly alluded to in \cite{boonkasame2012stability} (\S~3), it does not appear to be widely disseminated in the hydraulics literature.

\subsection{Link between complex characteristics and instability} \label{sec:link}

    \begin{figure}
	\centering		
	\includegraphics[width=\linewidth, trim=0mm 0mm 0mm 0mm, clip]{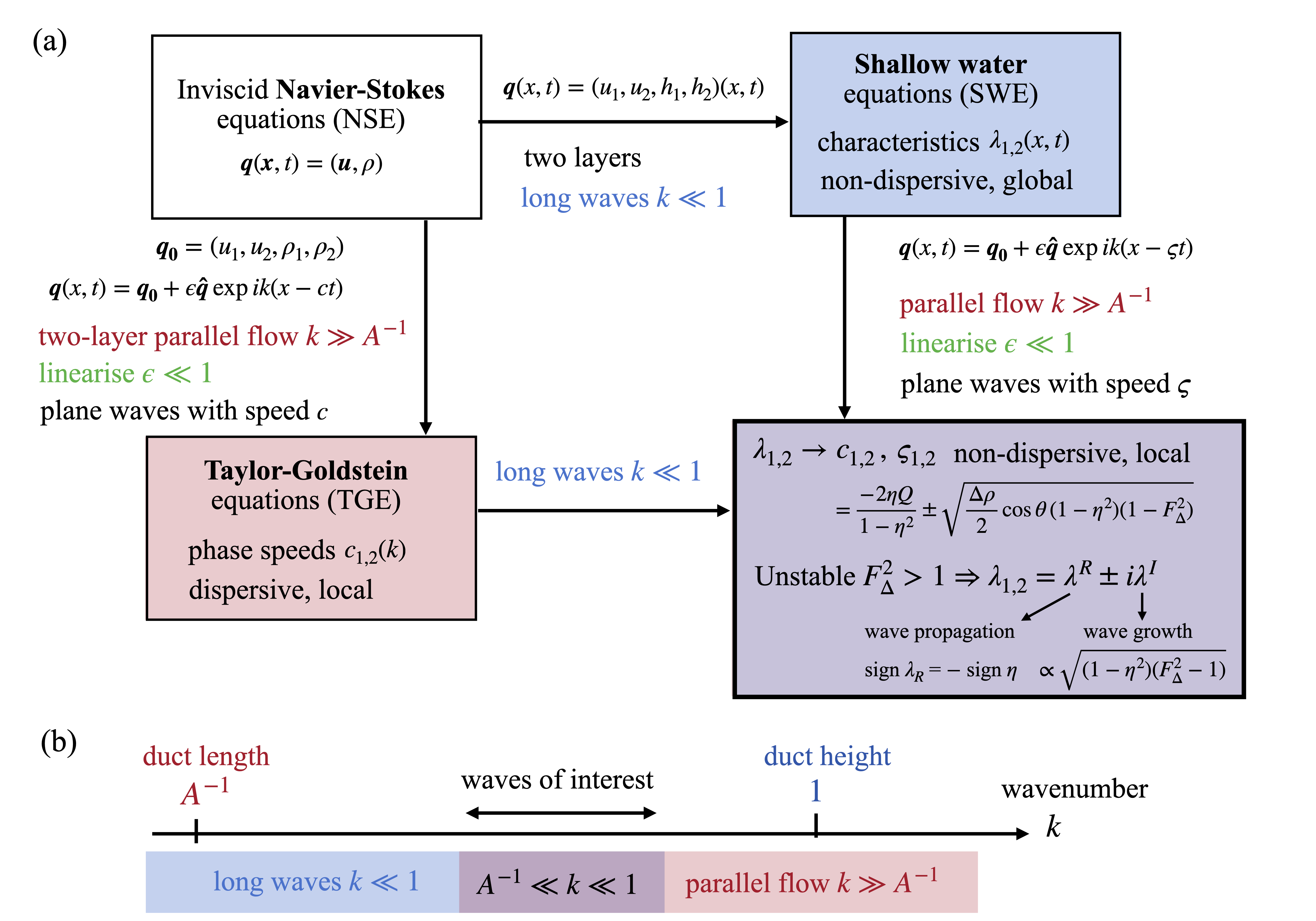}
 \caption{(a) Summary of the interpretation of  characteristics in \S\ref{sec:twoLayer_def} using linear theory, either by first linearising the NSE, and then taking the long-wave limit (TGE, \S\ref{discontinous_2L}) or vice versa  (SWE, \S\ref{sec:link}). Both approaches yield the same result for waves much shorter than the duct length but much longer than the duct height (range sketched in b).}
	\label{fig:schem_LSA}
    \end{figure}

This natural link between $\lambda_{1,2}$ and $c_{1,2}$ in the long-wave limit can be further understood by considering another limit.
It is possible to directly linearise the  SWE \eqref{eq:SWE_main}   to study the evolution of infinitesimal perturbations $\epsilon\tilde{\boldsymbol{q}}(x,t)$ ($0<\epsilon\ll 1$) on a parallel, steady base flow $\boldsymbol{q_0}=(u_1,u_2,h_1,h_2)$ akin to \eqref{eq:baseflow_TGE}. We perform a first-order Taylor expansion of the coefficient matrix from \eqref{eq:SWE_matrices} $\mathsfbi{A}(\boldsymbol{q}) = \mathsfbi{A}(\boldsymbol{q_0}+\epsilon \tilde{\boldsymbol{q}})$ and obtain
\begin{equation} \label{eq:taylor-expansion-swe}
		\mathsfbi{C}\frac{\p (\boldsymbol{q_0}+\epsilon\tilde{\boldsymbol{q}})}{\p t}+ \Big[ \mathsfbi{A}(\boldsymbol{q_0})+\epsilon \tilde{\boldsymbol{q}} \frac{\partial \mathsfbi{A}}{\partial \boldsymbol{q_0}}\Big] \Big[ \frac{\p \boldsymbol{q_0}}{\p x} +\epsilon \frac{\p \tilde{\boldsymbol{q}}}{\p x}\Big]=0,
	\end{equation}	
At order $\epsilon$ we have the linear, local SWE
\begin{equation} \label{eq:LSA_SWE_base}
		\mathsfbi{C}\frac{\p \tilde{\boldsymbol{q}}}{\p t}+\mathsfbi{A_0}\frac{\p \tilde{\boldsymbol{q}}}{\p x}=-\tilde{\boldsymbol{q}} \frac{\p \mathsfbi{A}}{\p \boldsymbol{q_0}} \underbrace{\frac{\p \boldsymbol{q_0}}{\p x}}_{= \, 0} ,
	\end{equation}	
where $\mathsfbi{A_0}\equiv \mathsfbi{A}(\boldsymbol{q_0})$ is the local constant coefficient matrix. Importantly, the right-hand side coming from the Taylor expansion, and acting as a forcing term, vanishes when we assume the base flow to vary slowly. Substituting the plane wave ansatz $\tilde{\boldsymbol{q}} =  \boldsymbol{\hat{q}}\exp{ik(x-\varsigma t)}$ gives the phase speed $\varsigma$ as the eigenvalue of the matrix pair $(\mathsfbi{A_0} ,\mathsfbi{C})$. The two distinct solutions $\varsigma _{1,2}$ of $\det(\mathsfbi{A_0}-\varsigma  \mathsfbi{C})=0$ then become identical to the local characteristics $\lambda_{1,2}$ (at a fixed $x,t$) derived in \eqref{2L_Char_exact}.

In other words, the nonlinear shallow water wave ($k\ll 1$) characteristics $\lambda_{1,2}$ can be interpreted as the linear shallow water waves phase speed $\varsigma_{1,2}$  propagating on a locally parallel base flow ($k\gg A^{-1}$), which are themselves the long wave, non-dispersive limit case of the Taylor-Goldstein two-layer phase speeds $c_{1,2}$. This result ultimately stems from the fact that, simply put, the linearisation and the long wave limit commute. In particular, the potential positive imaginary component of characteristics $\lambda^I>0$ can be interpreted as the exponential growth rate of such unstable waves satisfying $A^{-1} \ll k \ll 1$.

This interpretation is summarised in figure~\ref{fig:schem_LSA}. The long aspect ratio $A$ of SID (in this paper $A=30$), ensures that the range of waves $A^{-1} \ll k \ll 1$ exists, and therefore that this interpretation is useful, unlike in shorter geometries having $A\lesssim 10$. We also note in passing that this interpretation can be generalised to any number of layers greater than two.

\section{Two-layer hydraulics applied to DNS}
\label{sec:two-lay_hydr}

In this section, we use the modelling results of \S\ref{sec:char-instab} to understand the observations of hydraulic jumps and maximal exchange of \S\ref{sec:theory_2lay}. In  \S\ref{sec:application-FDelta}  we study the stability Froude number flagging long-wave instability, in \S\ref{sec:application-lambda-G2} we focus on the characteristics and composite Froude number to diagnose internal jumps and hydraulic control, and in \S\ref{sec:maxExchange} we explain the observed flow rate with notions of maximal exchange, viscous friction and mixing.

\subsection{Stability Froude number and instability} \label{sec:application-FDelta}

\begin{figure}
	\centering		
		\includegraphics[width=\linewidth, trim=4mm 0mm 3mm 0mm, clip]{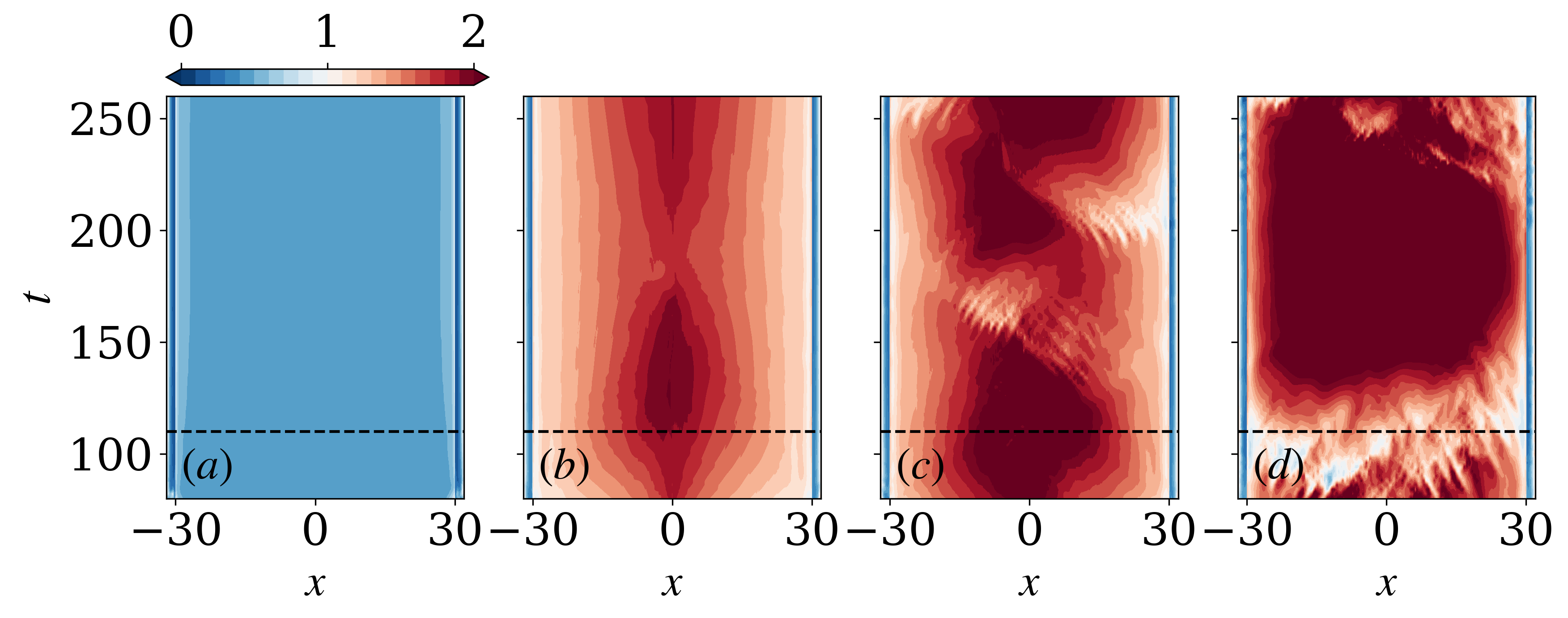}
	\caption{Spatio-temporal diagram ($x-t$) of the stability  Froude number $F_\Delta^2$  for the ($a$) L, ($b$) SW, ($c$) TW, and ($d$) I cases. Long-wave instability is predicted for $F_\Delta^2>1$ (red colour).}
	\label{fig:twolay_fdelta}
\end{figure}

In figure~\ref{fig:twolay_fdelta} we plot the spatio-temporal diagram of the stability Froude number $F_\Delta^2(x,t)$ given by \eqref{2L_Fr}  in our four datasets L, SW, TW, and I (all at $Re=650$) to diagnose any long wave instability.

The laminar flow (L) has $F_\Delta^2<1$ everywhere (in blue in figure~\ref{fig:twolay_fdelta}a), hence long waves are stable at $\theta=2^\circ$. 
In all other cases, $F_\Delta^2>1$  (in red in figure~\ref{fig:twolay_fdelta}) in most of the duct, hence long waves are unstable at $Re=650$ for $\theta> \theta_c$ where $\theta_c\in [2^\circ, 5^\circ]$. In the wave flows (SW and TW), the waves are most unstable (maximum $F_\Delta^2$, deep red) near the centre of the duct. 
In the I flow, long waves are very unstable ($F_\Delta^2\gg1$) throughout most of the duct. These $F^2_\Delta(x,t)$ diagrams correspond to the absence of a jump in the duct in the L flow (figure~\ref{fig:hu_TW}a) and  the existence of a jump in the SW, TW, and I flows (figure~\ref{fig:hu_TW}b-d), although the reasons for the jump remain to be explained.
These diagrams also show that as the tilt angle $\theta$ is modestly increased between $5^\circ$ and $8^\circ$, further changes take place as the flow becomes increasingly unstable to long waves.  Next, we delve deeper into the hydraulics analysis to understand jumps by focusing on the  characteristics in each flow.

\label{sec:twolay_DNS}

\subsection{Characteristics, composite Froude number and control} \label{sec:application-lambda-G2}

\begin{figure}
		\centering		
		\includegraphics[width=0.9\linewidth, trim=0mm 0mm 0mm 0mm, clip]{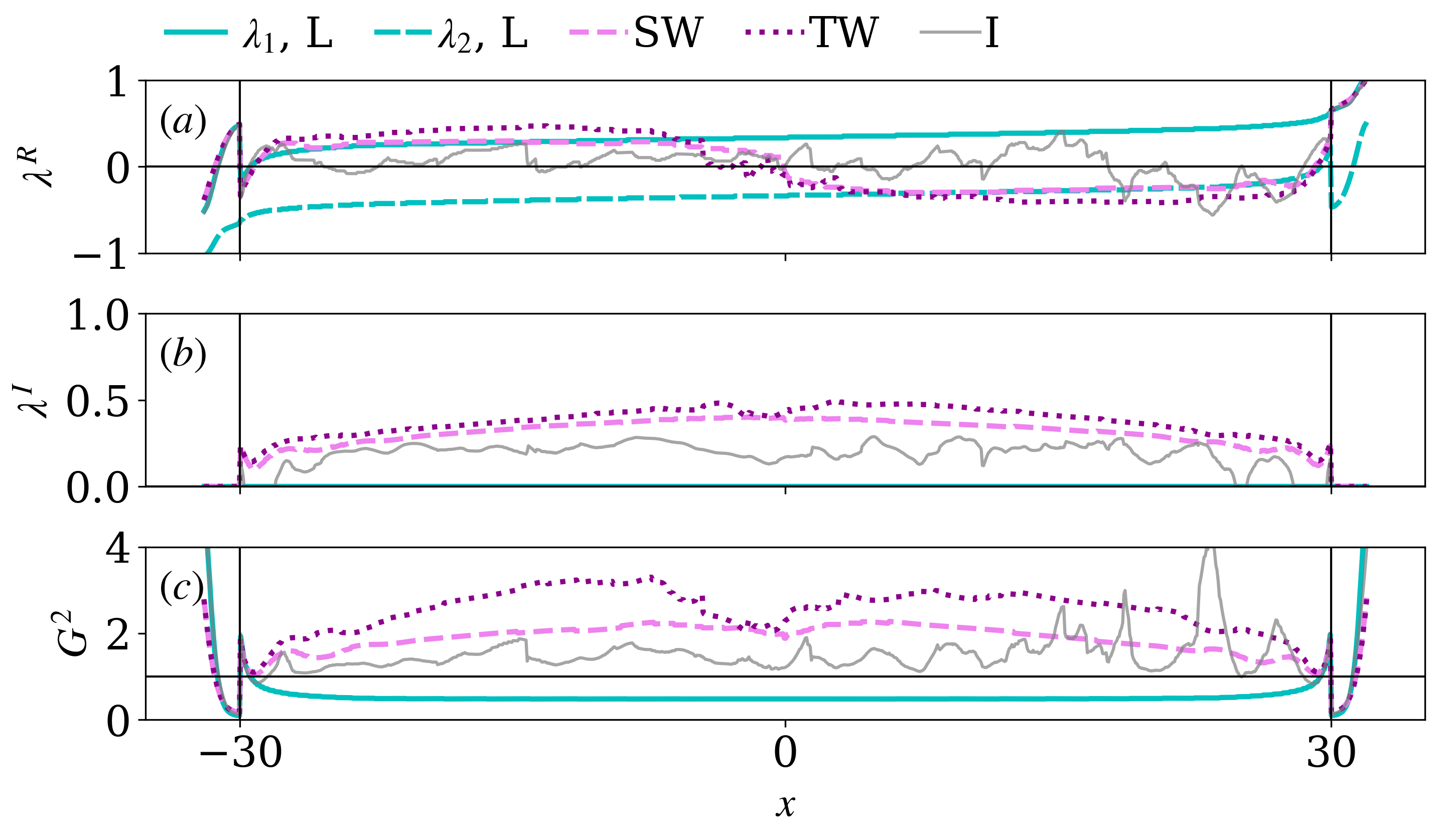}
		\caption{Characteristics (a) real part $\lambda^R$, (b) imaginary part $\lambda^I$ (only the positive values are shown),  and (c) composite Froude number $G^2$ of the L, SW, TW, I flows at $t=110$. Note that we also show data immediately outside the duct, up to $x=\pm 32$. 
  }

		\label{fig:twolay_eigval}
	\end{figure}

\begin{figure}
    	\centering	
    	\includegraphics[width=1.0\linewidth, trim=0mm 0mm 0mm 0mm, clip]{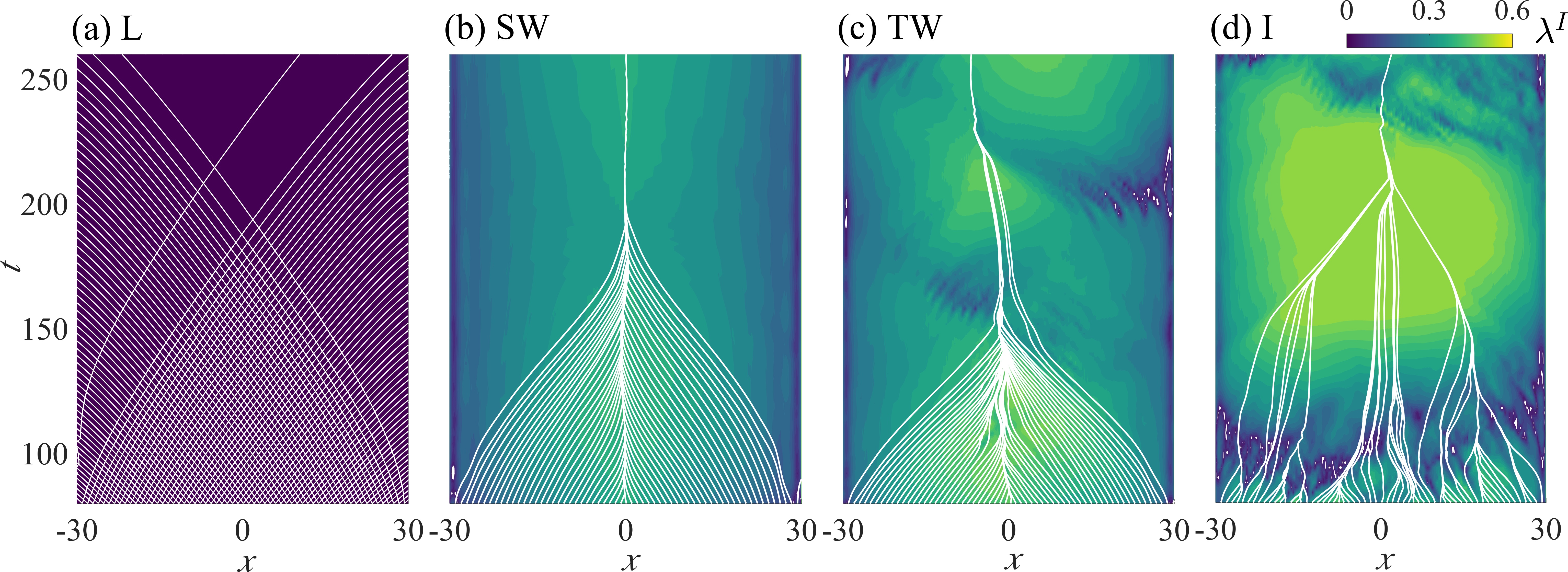}
        \caption{Characteristic curves $X(t)$  (in white) obtained by \eqref{eq:characteristic-plot} for (a) L, (b) SW, (c) TW, (d) I. The curves (in white) originate from 61 equally spaced positions between $x=-30$ and $30$. The colour contours show the growth rate $\lambda^I(x,t)$ (which is zero in a). Note that we only show data inside the duct up to $x=\pm 30$.}
        \label{fig:characteristics}
    \end{figure}

In figure~\ref{fig:twolay_eigval} we plot the characteristic velocities $\lambda_{1,2}(x)$ given by \eqref{2L_Char_exact} (real parts in panel a and imaginary parts in panel b) and the composite Froude number $G^2(x)$ given by \eqref{Gcomp_Lamab2} (panel c) at time $t=110$, to diagnose the propagation of information and criticality of the flow, respectively. In figure~\ref{fig:characteristics} we plot a set of discrete trajectories (white curves) by solving 
\begin{equation}\label{eq:characteristic-plot}
    \frac{dX}{dt} = \lambda^R(X,t) \ \ \text{for} \ t>80,
\end{equation}
and initialising $X(t=80)$ as 61 equidistant points along the duct $x\in [-30,30]$. Additionally, we plot the local growth rate $\lambda^I(x,t)$ in background colours (dark blue to yellow) for comparison.

	Figures~\ref{fig:twolay_eigval}(a) and \ref{fig:characteristics}(a) show that in the stable L flow,  $\lambda$ has two distinct real roots of opposite signs throughout most of the duct ($\lambda^R_1\lambda^R_2<0 \Leftrightarrow G^2<1$), allowing characteristics to cross.  Information propagates in both directions and the flow is subcritical inside the duct. The speed of propagation is of order $0.2-0.4$, i.e. significantly lower than the advective velocity 1. However, at the ends of the duct, the characteristics vanish locally  ($\lambda^R_{1} \lambda^R_2=0 \Leftrightarrow G^2=1$ at $|x|\approx 30$) signalling that long waves become stationary. \Cref{fig:twolay_eigval}(a)  shows that immediately outside the duct, information propagates only in one direction ($G^2>1$ at $|x|\gtrsim 30$), in fact, leftward on the left-hand side ($\lambda^R_{1},\lambda^R_2<0$) and rightward on the right-hand side ($\lambda^R_{1},\lambda^R_2>0$), i.e. always away from the duct. The ends of the duct, therefore, act as control points, in the sense that no information from the reservoirs can propagate into the duct. The existence of two such hydraulic control points, with their respective characteristics pointed outwards, means that the interior of the duct is `fully controlled' in the hydraulic sense, isolating the flow from hydrostatic disturbances within the reservoirs to either side. This is the first direct evidence that SID flows in the L regime are hydraulically controlled.
 
    In contrast, in the unstable SW, TW, and I cases, the roots are complex conjugates throughout most of the duct (figure~\ref{fig:twolay_eigval}(a,b)), a consequence of instability ($F_\Delta^2>1$), causing supercriticality ($G^2>1$, figure~\ref{fig:twolay_eigval}(c)). These unstable waves always move at the local convective velocity of the flow $\lambda^R=\bar\lambda(x,t)$, which we recall from \eqref{2L_Char_approx} is  non-zero if the interface is not at mid-depth $\eta(x,t)\neq 0$. 
 
    In the SW and TW cases, the unstable waves are initially carried rightward ($\bar\lambda>0$) throughout most of the left-hand side of the duct, and leftward ($\bar\lambda<0$)  throughout most of the right-hand side of the duct, as seen in figure~\ref{fig:characteristics}(b,c). Thus, all unstable waves are carried towards the centre ($x=0$) where their characteristics converge, creating the hydraulic jump observed in figure~\ref{fig:interface}.  The largest values of $\lambda^I=|\delta\lambda|$ are found in the region where the $\lambda^R=\bar\lambda$ convective components converge (see figure~\ref{fig:twolay_eigval}(b) and green-yellow shades in figure~\ref{fig:characteristics}(b,c)). Their growth rate is fast (${\lambda^I} \approx 0.2-0.5$) and slightly higher in TW than in SW. 
    Such a jump, bounded by supercritical regions on either side, is distinguished from the standard hydraulic jumps which make the flow transition from a supercritical to a subcritical state. It can be viewed as the limit of the length of the subcritical region tending to zero. The jumps in SW and TW can be called `undular jumps' because of their moderate `upstream' Froude numbers $F^2_i\approx G^2/2$ of each layer (between 1 and 2) and the small energy they dissipate, compared to direct, breaking hydraulic jumps. Some jumps in I are stronger, as evidenced by their locally higher $G^2$ values and their visibly higher dissipation. 
    
    This pattern of unstable characteristics converging toward the centre to form a jump can be summarised by $\textrm{sign} \, \lambda^R= - \textrm{sign} \, x$. This can be understood first by  $\textrm{sign} \, \lambda^R= \textrm{sign} \, \bar\lambda= - \textrm{sign} \, \eta$ from \eqref{2L_Char_approx}, i.e. the waves are carried at the local convective velocity, and second, by $\textrm{sign} \, \eta=  \textrm{sign} \, x$, i.e. the interface does not slope down as in L but is instead lowered on the left-hand side of the duct, and lifted on the right-hand side (central jump).

    The main difference between SW (stationary wave regime) and TW (travelling wave regime) lies in the behaviour of $\lambda_R$ near the jump around $x=0$ (figure~\ref{fig:twolay_eigval}(a)). While $\lambda_R(x)$ goes smoothly through zero in SW, it oscillates more in TW, suggesting that the location of the jump is prone to oscillations. This is confirmed by comparing the $x-t$ trajectory of the locus of the maximum  $F_\Delta^2$  in figure~\ref{fig:twolay_fdelta}(b,c) or the maximum $\lambda^I$ in figure~\ref{fig:characteristics}(b,c) as a proxy to the location of the jumps.

    Finally, we turn to the I flow, which exhibits more vigorous interfacial turbulence and wave instability (especially between $t=150-250$) and greater variability in $x$ and $t$ than SW and TW.  The characteristics in figure~\ref{fig:characteristics}(d) converge quickly to form a large number of local jumps around $t\approx 100$, which then organise into three main clusters: a central stationary cluster flanked by a left and a right cluster which themselves converge to form discrete jumps around $t\approx 160$ while being carried to the centre, eventually converging into a single jump  $t\gtrsim 200$. The convergence of characteristics tends to coincide with the maximum instability ($\lambda^I \approx 0.5$). During the more stable (transitional) period at $t=110$ the multiple sign reversals of  $\lambda^R(x)$  (figure~\ref{fig:twolay_eigval}(a)) hinder a straightforward interpretation of wave propagation along $x$. 
    
    This pattern by which unstable waves are carried in SW, TW and I flows differs so greatly from the classical picture of hydraulic control in L flow that it prompts the question: since information travels toward the duct centre, does it travel from the reservoirs into the duct, and is the flow still hydraulically controlled? 
    \Cref{fig:twolay_eigval} shows that close to the duct exits ($|x| \approx 28$), the composite Froude number $G^2$ (panel c) of all the cases becomes $1$. Meanwhile, immediately outside the duct  $30< |x| < 32$ the waves are stable ($\lambda_I= 0$, panel b), and both the curves of the real characteristics (panel a) and of the composite Froude number $G^2$ (panel c) closely match those in the L flow. In other words, the SW and TW flows also have a control point ($G^2=1$), flanked by narrow regions of subcriticality ($G^2<1$). We conclude that the flow within the duct  remains isolated from the reservoirs
    , and hence that it is also hydraulically controlled. This represents the first direct evidence that SID flows in the W and I regimes, in addition to having a (supercritical-to-supercritical)  jump in the centre of the duct, are also hydraulically controlled at the ends of the duct.

\subsection{Maximal exchange and critical flow rate}\label{sec:maxExchange}

In the previous section, we showed that all four flows cases (L, SW, TW and I) were hydraulically controlled in the sense that control points at the ends of the duct isolated the flow within the duct from processes in the reservoirs and prevented the flow inside the duct from reaching velocities exceeding a maximal volume flux $Q$. In this section, we seek to explain the differences in the value of this critical volume flux (and by extension mass flux) observed between the L, W, and I regimes in figure~\ref{fig:Qm}. 

In the stable L flow,  we find a time-averaged $Q\approx 0.31$  well below the absolute upper bound of $0.5$ for instability given by \eqref{2L_Fr_approx}. This is understood by the frictional hydraulic theory of \cite{gu_lawrence_2005}, subsequently adapted to SID in \cite{lefauve2020buoyancy} (their \S~5.2).  In short, the relatively low values of the Reynolds number in these low-tilt L flows mean that viscous friction at the duct walls and at the interface must be parameterised in the shallow water equations. This parameterisation allows a correct prediction of the sloping interface $\eta(x)$ (a consequence of viscous friction, i.e. loss of momentum along the flow of each layer), which in turn allows prediction of $Q$ by imposing the criticality condition at the ends of the duct $G^2(x=\pm A)=1$. Simply speaking, the lower $Re$ and the longer the duct aspect ratio $A$, the more friction occurs along the duct, the more offset the interface $|\eta|$ becomes at the ends of the duct, and the lower the volume flux $Q$ becomes to satisfy $G^2(|\eta|,Q)=1$ (since $G^2$ is an increasing function of both $|\eta|$ and $Q$).

In the unstable SW, TW and I flows, despite the existence of viscous friction, we find a remarkably consistent time-averaged $Q\approx 0.51-0.53$, slightly above the critical $Q_c=0.5$ upper bound for frictionless two-layer hydraulics and a flat interface $\eta=0$ (figure~\ref{fig:twolay_theo}). These values require a different explanation. Although the Reynolds number of SW, TW, and I are identical to L, their larger tilt angle $\theta$ pushes these flows beyond the instability threshold $F_\Delta^2=1$, corresponding to the transition between `lazy' and `forced' flows  \citep{lefauve2019regime}. However, $Q$ does not  continue to increase with $\theta$. Rather, beyond the instability threshold, \eqref{2L_Fr_approx} suggests that, in the centre of the duct (where $\eta\approx0$), $Q=0.5 \sqrt{(\Delta \rho)/2 \cos\theta}F_\Delta$, i.e. a linear increase with the stability Froude number. We deduce that, since $Q$ never greatly exceeds 0.5 (the value reached the instability threshold), the subsequent increase in  $F_\Delta>1$ (indirectly caused by the forcing $\propto\ sin \theta$ in the DNS) must be compensated by a decrease in $\Delta \rho/2\propto 1/F^2_\Delta$, i.e. by increased mixing. The data show that the average $\Delta \rho/2$ indeed decreases from 0.79 in SW, to 0.76 in TW, to 0.68 in I. This mixing in turn explains why $Q_m$ (roughly $\approx (\Delta\rho/2)Q$) stays robustly below 0.5 in SID, and indeed decreases from the W to the I regime (figure~\ref{fig:Qm}).

 \section{Applicability of unstable hydraulics}
    \label{sec:twoLay_LSA_TG}

In this section, we study the applicability of the previous results to problems not usually considered in two-layer hydraulics. In \S\ref{sec:top-bottom-walls} we study the transition between long waves (the propagation of which is identical to the local characteristics of the SWE) and shorter Kelvin-Helmholtz waves (only predicted by the TGE). In \S\ref{sec:influence-of-Pr} we study the waves diagnosed from DNS run at different values of the Prandtl number to investigate their indirect dependence on scalar diffusion and the thickness of the density interface. In \S\ref{sec:continuous_flow_profiles} we study how the growth of long waves is impacted by smooth, diffuse (i.e. not strictly two-layer) density and velocity profiles, which are expected in all real-world flows (having a finite $\Rey$ and $\Pran$).

\subsection{Long vs short waves}\label{sec:top-bottom-walls}

\begin{figure}
    \centering		
        \includegraphics[width=0.9\linewidth, trim=0mm 0mm 0mm 0mm, clip]{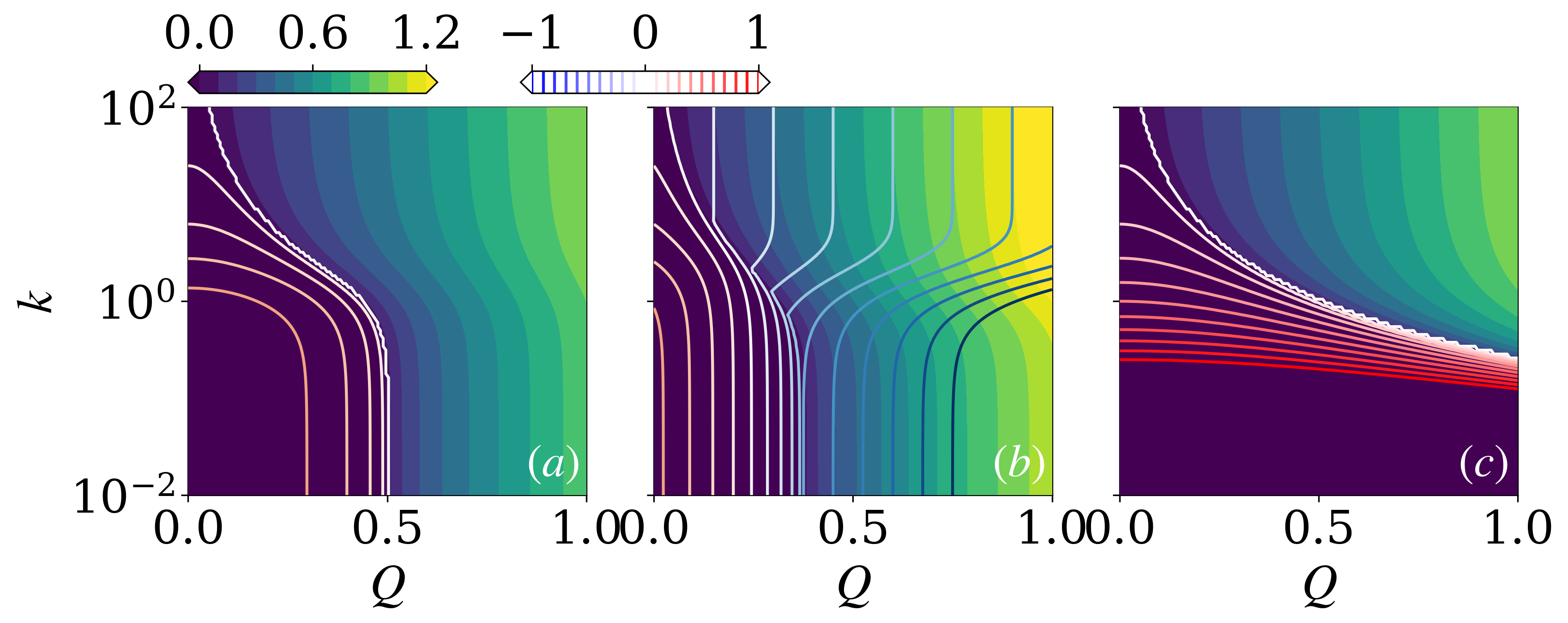}
        \caption{Dispersion relation of all (long and short) inviscid two-layer waves: growth rate  (colours) and phase speed (contours) solutions  of the TGE varying with wavenumber $k$ and volume flux $Q$. (a)  Symmetric interface $\eta=0$. (b) Asymmetric interface $\eta=-0.5$. (c) Symmetric interface but without solid top and bottom boundaries at $z=\pm 1$, in which case the long waves of SWE disappear. 
        }
    \label{fig:TG_2L_Q}
    \end{figure}

    \Cref{fig:TG_2L_Q} shows the dispersion relation from the  TGE \eqref{lambda_2L_TG_mt} with the phase speed $\Re(c_1) = c_1^R$   (blue to red contours) and growth rate $\Im(c_1) = c_1^I$ (colour map with deep blue being stable) as functions of the wavenumber $k$ (vertical axis) and the volume flux $Q$ (horizontal axis). We compare a symmetric interface ($\eta=0$, panel a), an asymmetric interface ($\eta=-0.5$, panel b) and a case with a symmetric interface but without solid top and bottom boundaries (panel c), whose  dispersion relation \eqref{eq:TG-disp-rel-no-walls} is derived analytically in \S\ref{sec:TGderiv-nowalls}.
    
    We recall that for $k \ll 1$ the phase speed $c^R$ and growth rate $c^I$ of TGE become identical to $\lambda^R$ and $\lambda^I$ of SWE, respectively. In this case the $k=10^{-2}$ data of figure~\ref{fig:TG_2L_Q}(a,b) become indistinguishable from those plotted in figure~\ref{fig:twolay_theo}(b,c), respectively.
   
        \begin{figure}
    \centering		
    \includegraphics[width=0.65\linewidth, trim=0mm 0mm 0mm 0mm, clip]{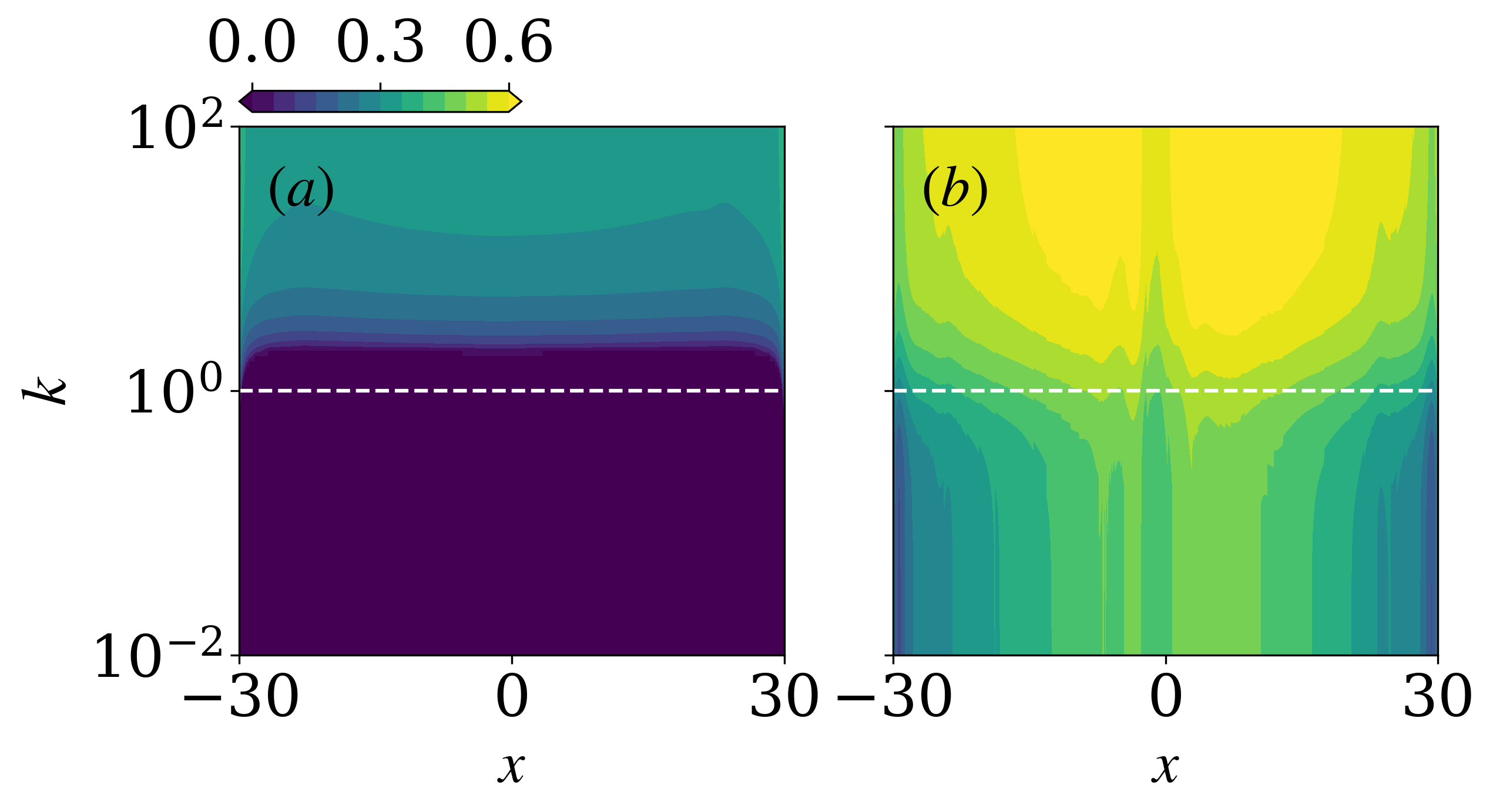}
    \caption{Dispersion relation (growth rate only) predicted by the TGE two-layer model applied to the DNS  (a) L and (b) TW flows. The dashed line at $k=1$ represents the boundary between long and short waves. Short waves are predicted to be most unstable by this inviscid model but appear relatively stable in reality.}
    \label{fig:TG_dns}
    \end{figure}

From figure~\ref{fig:TG_2L_Q}(a-b) we recover the results from previous sections. First, long waves are non-dispersive (the phase speed contours do not depend on $k$). Second, they become unstable (lighter shades of blue and green) above a critical volume flux $Q> Q_c$, equal to $0.5$ for a symmetric interface (panel a) and lower than $0.5$ for an asymmetric interface (panel b). Third, for a symmetric interface, all unstable waves (long and short) are stationary (absence of contours), but all stable waves are travelling (presence of contours). For an asymmetric interface, even unstable waves are travelling in the reference frame of the duct, as we explained in \S\ref{sec:application-lambda-G2}. Fourth, the transition between short and long waves is smooth, i.e. there is a continuity between the long shallow water waves controlling the hydraulics of two-layer flows and the short KH waves.
    
Panels (a,b) of figure~\ref{fig:TG_2L_Q} also give new results. First, short waves ($k \not\ll 1$) become unstable at smaller values of $Q$ compared to the long wave threshold $Q_c$. This transition to short waves becomes noticeable from $k\gtrsim 10^{-0.5}\approx 0.3$ and is clear for $k>1$. The shortest waves shown here ($k=10^2$) are predicted to become unstable above a very small volume flux $Q\gtrsim 0.1$. However, we note that this threshold would be closer to the long-wave $Q_c$ if we included viscosity in the TGE, as viscosity would significantly damp the growth of short waves. These panels show that for a given value of $Q$, the growth rate increases monotonically with $k$ (i.e. the shortest waves are the most unstable). As is often the case, this `ultraviolet catastrophe' would be regularised by viscosity, with the possible existence of a maximum growth rate at an intermediate $k$ for intermediate values of $\Rey$. 

In figure~\ref{fig:TG_2L_Q}(c), the absence of solid walls does not affect short waves (the colours and contours at $k\gtrsim 3$ are identical to panel a), because they do not `feel' the presence of the walls. However, the absence of solid walls precludes the existence of long waves $k\lesssim 0.3$, because this setup (despite being bounded at $z= \pm 1$) approximates layers of infinite depth, compared to which all waves are `short'. In other words, this analysis explicitly shows that the presence of solid walls in SID are crucial to explain the leading order dynamics in the DNS by allowing  long-wave instability. Adding walls (panel a) creates the long waves on which hydraulic effects rely, an long waves transition smoothly into short waves as $k$ increases.

\Cref{fig:TG_dns} shows the linear growth rate $c^I$  obtained by substituting into the TGE dispersion relation \eqref{lambda_2L_TG_mt} (function of $k$) the two-layer properties  (as functions of $x$) extracted from the DNS. Diagnostics are shown for the L flow (panel a) and the TW flow (panel b) at time $t=110$.  The TW flow is, unlike the L flow, unstable to long waves $k\ll 1$, with the maximal growth rate found near the centre of the duct, as previously seen in figure~\ref{fig:twolay_eigval}(b) as we know from \eqref{eq:limit} that $c^I(k\ll 1)\longrightarrow \lambda^I$. However, we also find that the L flow appears mildly unstable to short waves (especially very short waves $k\gtrsim 10$), and the TW flow appears even more strongly unstable to them. However, we know that in the DNS the L flow is visibly stable and does not have observable interfacial waves, while the TW flow is primarily unstable to long waves, whereas short waves play a more minor role. This suggests that, at least for the present values of $\Rey=650$ and $\Pran=7$, the growth of short waves is sufficiently damped by viscosity, mass diffusion and/or other effects not taken into account in this inviscid two-layer model.

\subsection{Low vs high Prandtl numbers} \label{sec:influence-of-Pr}

Although $\Pran$ does not appear explicitly in the SWE, the DNS dynamics from which the two-layer properties are extracted certainly depend on $\Pran$. In applications, three typical values are of particular interest: $\Pran\approx 1$ (representative of temperature stratification in air), $\Pran\approx 7$  (representative of temperature stratification in water, as studied in this paper) and $\Pran\approx 700$ (representative of salt stratification in water).

\begin{figure}
	\centering		
	\includegraphics[width=0.9\linewidth, trim=0mm 0mm 0mm 0mm, clip]{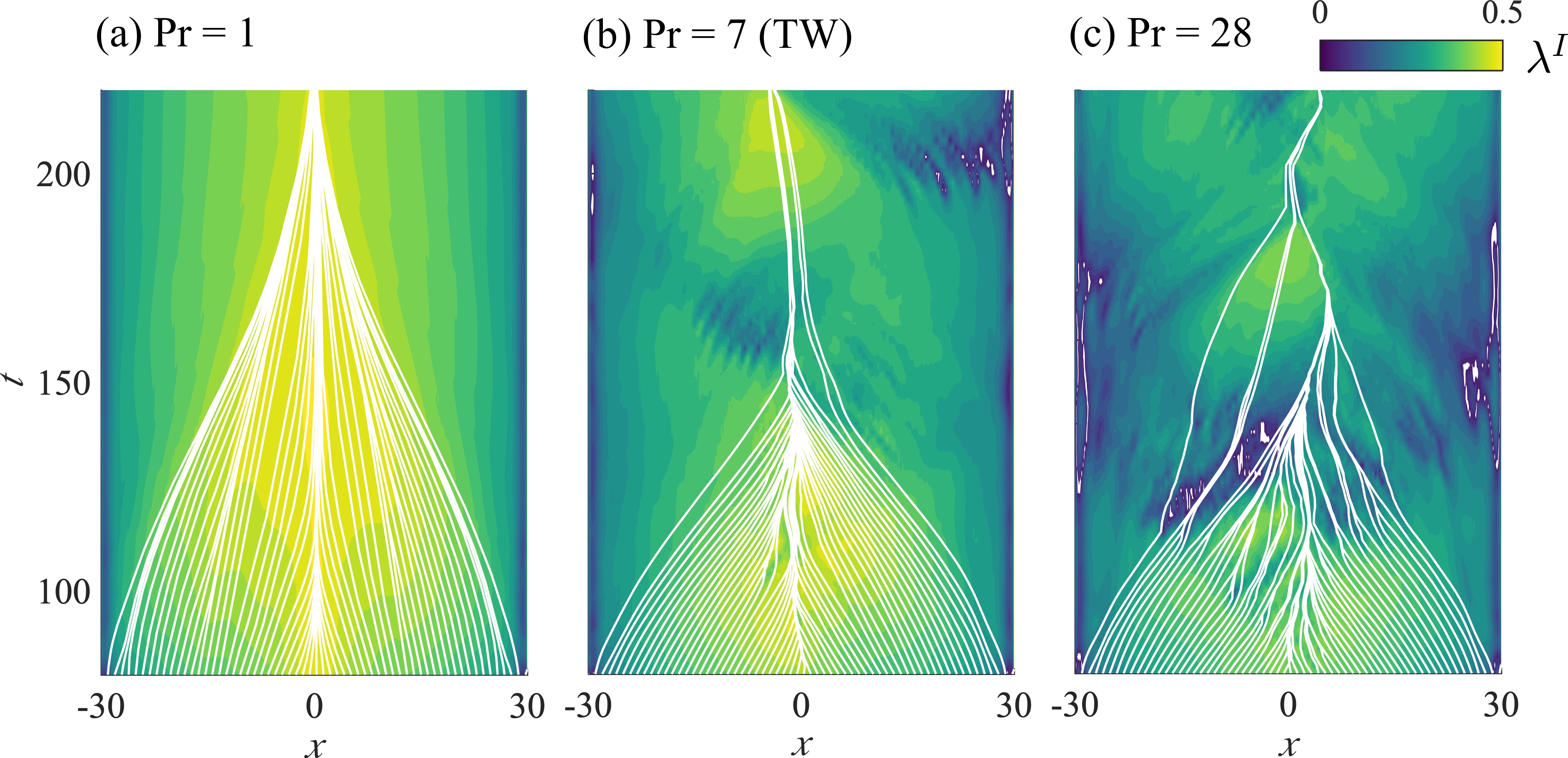}
	\caption{Characteristic curves $X(t)$ and growth rate $\lambda^I$ (colours) from DNS data at $\Rey=650$ $\theta=6^\circ$ and (a) $\Pran=1$, (b) $\Pran=7$ (TW data), (c) $\Pran=28$. Note that for easier comparison, (b) is identical to figure~\ref{fig:characteristics}(c), though we only show data for $t\in[80,220]$. Also note the reduced colour map maximum $\lambda^I$ (0.5 in this figure, compared to 0.6 in figure~\ref{fig:characteristics}) }
	\label{fig:twolay_char-Pr}
    \end{figure}

To study the impacts of diffusion, we carried out two additional DNS with parameters identical to the TW flow (with $\Pran=7$) at $\Pran=1$ and $\Pran=28$ (the latter requiring a much higher spatial resolution, hindering the study of higher $\Pran$).  In figure~\ref{fig:twolay_char-Pr} we compare the characteristics curves $X(t)$ and growth rates $\lambda^I$ at these three different values of $\Pran$ using the same visualisation as figure~\ref{fig:characteristics} (where the $\Pran=7$ data was already shown as TW). We find that curves from the $\Pran=1$ flow initially converge into a central jump and a small number of peripheral jumps, which eventually merge with the central jump. This pattern resembles that of the more stable SW flow from figure~\ref{fig:characteristics}(b), except that it has a higher growth rate than TW. The curves from the $\Pran=28$ flow converge into a larger number of intermediate, travelling jumps before merging into a single jump. This pattern resembles that of the more unstable I flow from figure~\ref{fig:characteristics}(d), except that it has a smaller growth rate. than TW. 

 \begin{figure}
	\centering		
	\includegraphics[width=.85\linewidth, trim=0mm 0mm 0mm 0mm, clip]{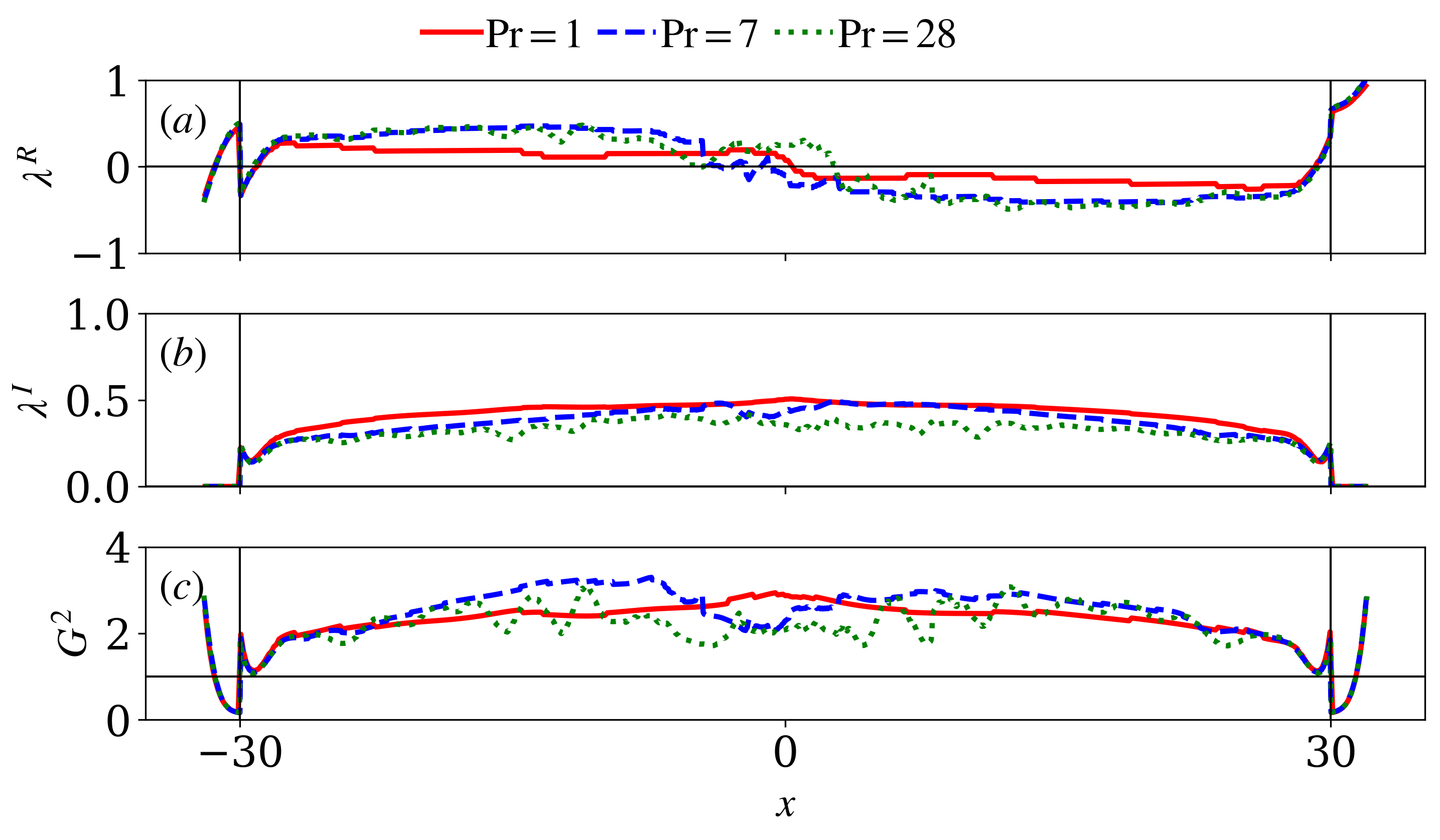}
	\caption{Characteristics (a) $\lambda^R$, (b) $\lambda^I>0$,  and (c) composite Froude number $G^2$ of TW flows at $t=110$. We compare three different Prandtl numbers  $\Pran=1, \ 7,\, 28$ as in figure~\ref{fig:twolay_char-Pr}. }
	\label{fig:twolay_eigval_Pr}
    \end{figure}

In figure~\ref{fig:twolay_eigval_Pr} we compare the characteristics $\lambda^R(x)$ (panel a) and $\lambda^I(x)$ (panel b) as well as the composite Froude number $G^2(x)$ at $t=110$ (the $\Pran=7$ data was already shown in figure~\ref{fig:twolay_eigval}). All curves (solid red, dashed blue, and dotted green) have essentially the same qualitative features described earlier in figure~\ref{fig:twolay_eigval}. However, as noted in figure~\ref{fig:twolay_char-Pr}, the growth rate appears to decrease slightly with $\Pran$. 

The synoptic features of the flow governed by long waves, therefore, appear relatively unaffected by $\Pran$. This can be rationalised by the fact that $\Pran$ will primarily influence the thickness of the density interface separating the two layers, rather than its location $\eta$ (the locus of $\rho=0$) or the speed of the flow ($Q$), which are the two key model variables in the SWE. 

We expect short waves to be more strongly influenced by a decreasing thickness of the density interface with increasing $\Pran$, and vice versa. However, the short waves observed in our DNS at $\Pran=28$ and in experiments at $Pr=700$ are Holmboe waves, not the KH waves supported by our two-layer model. Unlike the KH instability caused by a single vortex sheet, the Holmboe instability is caused by the resonance of vorticity waves (e.g. on the edges of a diffuse velocity interface) with a non-collocated gravity wave (on a sharper density interface)  \citep{carpenter_instability_2011}. \cite{lefauve2018structure} performed a linear stability analysis on the experimentally measured mean flow (at $\Rey = 440$, $\Pran=700$), including viscosity and scalar diffusion. They found that intermediate $1\lesssim k \lesssim 2$ Holmboe waves were most unstable (see their figure 6a). However, tackling Holmboe waves -- and their presumed coexistence with the long waves governing hydraulic processes at high $\Pran$ SID -- would require a three-layer model (for velocity) mixed with a two-layer model (for density).

 \subsection{Sharp vs smooth two-layer flow profiles} \label{sec:continuous_flow_profiles}

\begin{figure}
    \centering		
       \includegraphics[width=0.95\linewidth, trim=0mm 0mm 0mm 0mm, clip]{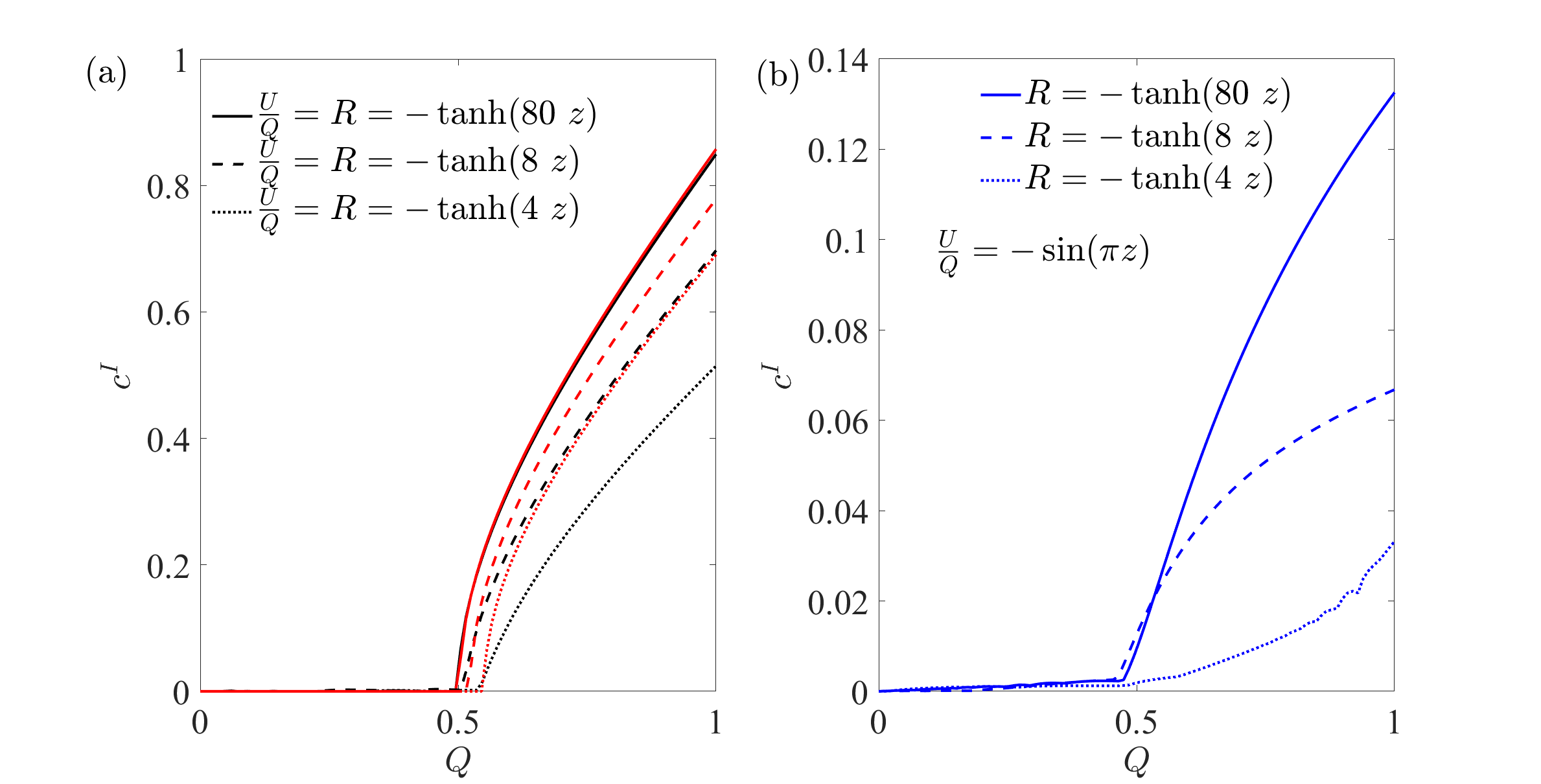}
        \caption{Growth rate $c^I$ of long waves on smooth velocity and density profiles, obtained by a numerical solution of the TGE \eqref{TG_eq} as the volume flux $Q$ is increased (black curves). (a) Hyperbolic-tangent profiles for velocity $\mathcal{U}(z)$ and density $\mathcal{R}(z)$ profiles of increasing interface thickness. (b) Sinusoidal profiles for velocity $\mathcal{U}(z)$ (and same $\mathcal{R}(z)$ as in (a)). The growth is always slower than it would be using the sharp two-layer analytical solution  \eqref{lambda_2L_TG_mt} (red curves).}

    \label{fig:TG_2L_long_Q}
    \end{figure}

Finally, we study the influence of smooth density and velocity profiles $\mathcal{U}(z)$ and $\mathcal{R}(z)$ on the growth rate of long waves. To do so, we solve the eigenvalue problem from the Taylor-Goldstein equation \eqref{TG_eq}  before the two-layer base flow ansatz \eqref{eq:baseflow_TGE}. Numerical solutions for the growth rate $c^I$   are shown in figure~\ref{fig:TG_2L_long_Q} as functions of $Q$. In panel a, we show the results for hyperbolic-tangent $\mathcal{U}(z)/Q=\mathcal{R}(z)=\tanh \, z/\delta$ where the interface thickness is progressively decreased from $1/80$ (almost exactly two layers, solid lines) to $1/8$ (dashed lines) to $1/4$ (thicker interface, dotted lines). We compare these `smooth tanh' growth rates (in black) to the equivalent `sharp two-layer' growth rates (in red) obtained from the analytical dispersion relation \eqref{lambda_2L_TG_mt} by layer-averaging the $\tanh$ profiles (in which case $c^I=\lambda^I$). In panel b, we keep the same density profiles but use $\mathcal{U}(z)/Q=-\sin \, \pi z$, which is a good approximation of the mean velocity at these relatively low values of $\Rey=O(10^2-10^3)$.

Both panels a and b show that the $Q_c=0.5$ threshold for long-wave instability is virtually unchanged by smooth profiles, with only a slight increase of a few percent for the thickest interface $\delta=1/4$. This result supports the relevance of two-layer hydraulics even in `real-world' flows which depart significantly from the two-layer model. 

However, the `smooth tanh' growth rates are always lower than the corresponding `sharp two-layer' growth rates. The thicker the interface, the lower the `smooth tanh' growth rate. Comparing the vertical scale in panels a and b, we conclude that the sinusoidal velocity profile is more stable (by approximately a factor of 10) than the tanh profile. The combination of a sinusoidal velocity and a diffuse velocity interface (dotted blue line in panel b) yields the slowest growth as $Q$ increases. As such profiles are a better approximation of the mean flow of the DNS at low $\Pran$ (e.g. $\Pran=1$) than sharp two-layer profiles, these results warn us not to interpret the large growth rates $\lambda^I=O(0.1)$ found  in this paper too literally. 

In other words, although the qualitative predictions of two-layer hydraulics are robust (in particular the long-wave instability threshold $Q_c$)  when the underlying  data is not exactly two-layer, the quantitative growth rates predictions are over-estimated when the interface is diffuse, as in low-$\Pran$ flows.

\section{Conclusions}\label{sec:conclusion}

In this paper, we employed a two-layer averaging procedure to extract a reduced-order representation of
four direct numerical simulations (DNS) datasets 
in the stratified inclined duct (SID) at $\Rey=650$ and $\Pran=7$ (with two supplementary datasets at $\Pran=1$ and $\Pran=28$. This two-layer representation  revealed in \S\ref{sec:theory_2lay} that the flow is stable in the laminar regime (L, tilt angle $\theta=2^\circ$), but develops an internal hydraulic jump (discontinuity in the layer properties) in the centre of the duct in the stationary wave regime (SW, $\theta=5^\circ$). This jump  moves around in the travelling wave (TW, $\theta=6^\circ$) regime, and causes further disorganised wave breaking in the intermittently turbulent (I, $\theta=8^\circ$) regime.

\subsection{Modelling results}

To understand these findings, in \S\ref{sec:char-instab} we adapted to SID DNS the well-known inviscid Boussinesq shallow water equations (SWE) governing the nonlinear evolution of long waves ($k\ll 1$) at a sharp density interface. The SWE predict that information propagates along a pair of trajectories, $\lambda_{1,2}$, that arise from the solution of a generalised eigenvalue problem and depend on the local ($x$) and instantaneous ($t$) state of the two-layer representation. The solutions can be written in the form $\lambda_{1,2}=\bar\lambda\pm\delta\lambda$, where the convective velocity $\bar\lambda$ is always real but the phase speed $\delta\lambda$ may be either real or imaginary. When $\lambda_{1,2}$ are real ($\delta\lambda\in\mathbb{R}$), they represent the propagation of two (neutrally stable) kinematic waves where $\bar\lambda$ can be interpreted as a convective velocity and $\delta\lambda$ as the phase speed of waves relative to $\bar\lambda$. The respective signs of $\lambda_{1,2}$ determine the direction of information propagation and whether the flow is subcritical (composite Froude number $G^2<1$; product $\lambda_1\lambda_2<0$) with information propagating in both directions, or supercritical ($G^2>1$; $\lambda_1\lambda_2>0$) with information propagating only in the direction given by the sign of $\bar\lambda$. When $\lambda_{1,2}$ are complex $\lambda_{1,2}=\lambda^R \pm i\lambda^I=\bar\lambda\pm i|\delta\lambda|$, the real part still represents a convective velocity that carries information while the positive imaginary part indicates that the flow is unstable. Although in this unstable case the SWE are no longer hyperbolic, the flow may be viewed as supercritical ($G^2>1$) in the sense that information is propagated only in the direction given by $\bar\lambda$. 

To interpret the unstable SWE, we compared the characteristics with the dispersion relation from the inviscid Taylor-Goldstein equation (TGE) governing the linear normal-mode stability of a two-layer base flow. We showed that the global nonlinear characteristics $\lambda(x,t)$ defined on the non-parallel base flow and sloping interface of the SWE could be interpreted locally as the phase speed and growth rate of linear waves in the  long-wave limit (i.e. $k\ll 1$), propagating on a base flow that is locally assumed parallel. Importantly,  this interpretation is only valid for waves that are much shorter than the duct length $2A$ (i.e. $k\gg A^{-1}$). It provides a local, linear stability interpretation for unstable two-layer wave characteristics satisfying $A^{-1} \ll k \ll 1$, which is a relevant range in long ducts ($A^{-1}\ll 1$). The dispersion relation for the dispersive TGE waves $c(k)$ tend to the non-dispersive SWE characteristics $\lambda$ as $k\ll 1$, but they also allow us to explore the smooth transition to shorter ($k\not\ll 1$), non-hydrostatic Kelvin-Helmholtz (KH) waves. 

\subsection{Physical results}

Applying this two-layer hydraulics and instability framework to the two-layer-averaged DNS datasets  yielded the main physical results of this paper in \S\ref{sec:two-lay_hydr}. We provided the first direct evidence that SID flows are, in all regimes (L, SW, TW and I), hydraulically controlled at the ends of the duct and thus in a state of maximal exchange. At these control points, the flow is locally supercritical; thus information from the reservoirs cannot enter the duct and influence the flow within it. In the SW, TW and I regime, the flow in the duct is always unstable to long waves ($F_\Delta^2>1$) and thus supercritical ($G^2>1$), explaining the existence of an undular jump within the duct, as a consequence of characteristic trajectories converging to a single point. In the I regime, multiple local jumps gradually merge into clusters and eventually into a single, stronger jump, resulting in greater instability. 

The emergence of unstable, supercritical flow in SID beyond a certain tilt angle is rationalised by the fact that gravitational forcing continuously provides a surplus of kinetic energy which must be dissipated. From a hydraulics perspective, the required dissipation in a supercritical flow (i.e. having a surplus of kinetic energy compared to potential energy) must be associated with an decrease in kinetic energy and an increase in potential energy, hence a thickening of both layers downstream of the jump. The physical insight of energy surplus and dissipation dates back to \cite{meyer2014stratified}. It was later formalised by \cite{lefauve2019regime} and \cite{lefauve2020buoyancy}, who showed using frictional two-layer hydraulics with a tilt $\theta$ that the mid-duct interfacial slope obeyed $\eta'(0)\propto \theta-F$, where $F$ represents viscous friction along the duct. The transition from subcritical to supercritical flow that we identified corresponds to  the transition from `lazy' to `forced' flows, which they identified based on the relative importance of the tilt $\theta$ and the duct geometric angle $\alpha=\tan^{-1}A^{-1}\approx 2^\circ$ in this paper. `Lazy' flows are characterised by  $\theta<\alpha$, and an interface gently  sloping down.  `Forced' flows are characterised by $\theta>\alpha$, and a relatively flat interface all along the duct. In forced flows, the tendency of $\theta$ to tilt up the density interface exceeds the tendency of frictional losses $F$ to tilt it down, thus $\eta'(0)>0$, which causes central jumps.

Next, we rationalised the values of the volume flux in all regimes. The value $Q\approx 0.3$ in the L regime (lazy flow) is explained by the offset of the interface  at the ends of the duct where control ($G^2=1$) takes place, recalling that the sloping interface is caused by viscous friction along the duct. The robust values $Q\approx 0.5$ in the unstable SW, TW, and I regimes (forced flows) are all surprisingly close to the instability threshold $Q_c=0.5$ for a symmetric interface.  Increasing instability from SW to TW to I as the tilt angle $\theta$ is increased (which theory predicts should increase $Q$ above 0.5) appears balanced by increasing mixing in the layers (which decreases $Q$). This explains why the maximal exchange threshold $Q=0.5$ predicted by inviscid long wave theory remains a remarkably robust feature of SID flows, even under turbulence.

Using the TGE analysis provided further physical insight into the applicability of unstable hydraulics in \S\ref{sec:twoLay_LSA_TG}. We showed that short inviscid KH waves are always predicted to be more linearly unstable than long waves, despite the fact that long waves cause the internal hydraulic jumps observed in SW, TW and I and appear to dominate the dynamics of these SID flows. 
We explained this paradox by the neglect of viscosity in TGE, which would damp the shortest waves. We also showed that DNS at lower $\Pran=1$ or higher $\Pran=28$ showed qualitatively (but not quantitatively) similar two-layer long wave hydraulics to $\Pran=7$. We also highlighted that experimental observations at $\Pran=700$ of the simultaneous existence of unstable long waves (causing an internal jump and supercritical flow) with short finite-amplitude Holmboe waves could not be explained by the two-layer model, because it does not support Holmboe waves. Finally, we showed that the predictions of long wave instability (especially the threshold $Q_c=0.5$) were robust even in diffuse two-layer flows having a thick interface.

\subsection{Outlook}

These key `hydraulic' features of SID flows, explaining the emergence of waves, increasingly supercritical jumps, and ultimately turbulence, result from long wave instability which (tautologically) relies on the existence of top and bottom solid boundaries confining the flow in a long, tilted duct. This `long-wave' pathway to turbulence in SID appears a priori to differ from the classical `short-wave' KH pathway in an unbounded stratified shear layer (see e.g. \cite{caulfield_anatomy_2000,mashayek_peltier_2012}), often used as a paradigm for ocean mixing. Further work is needed to clarify the relative importance of long and short waves, and within short waves, of Kelvin-Helmholtz (two-layer) waves and Holmboe (three-layer) waves, and how they contribute to the transition to turbulence under varying $\theta, \Rey, \Pran$.

The current formulation of the two-layer SWE does not account for the mixing layer that develops and appears to be important beyond the wave regime. The omission of a mixed layer may reduce the accuracy when investigating detailed spatial features of the flow, such as the localization of unstable wave regions in the centre of the duct in TW and SW. This model is able to predict the formation of shocks and provide clues to the long-length-scale dynamics and how they govern the synoptic features of the flow, but further work is needed to accurately represent the non-hydrostatic processes of turbulent mixing itself.

\vspace{0.3cm}

\textbf{Acknowledgments}\par
We acknowledge support from the European Research Council (ERC) under the European Union's Horizon 2020 research and innovation Grant No 742480 `Stratified Turbulence And Mixing Processes' (STAMP). Parts of the simulations were carried out with resources from Compute/Calcul Canada. A. L. is supported by a Leverhulme Trust Early Career Fellowship. For the purpose of open access, the authors have applied a Creative Commons Attribution (CC BY) licence to any Author Accepted Manuscript version arising from this submission.

\vspace{0.3cm}

{\textbf{Declaration of interests}} \par
The authors report no conflict of interest.
\vspace{0.3cm}

\appendix

\section{Hydrostatic and non-hydrostatic pressure gradients}\label{sec:dns_pre}

The assumptions behind the SWE require flows to be dominantly hydrostatic. To validate this assumption, we explicitly decompose the non-dimensional pressure into a hydrostatic component defined by
\begin{eqnarray}
p_h(x,z,t)=-Ri \cos{\theta}   \int_{-1}^z \langle \rho \rangle_y(x,\xi,t)  \ \textrm{d}\xi,  
\end{eqnarray}
and the remaining non-hydrostatic (but still spanwise averaged) component 
\begin{eqnarray}
p_{nh}(x,z,t)= \langle p \rangle_y-p_h.
\end{eqnarray}

\begin{figure}
	\centering		

	\includegraphics[width=.5\linewidth, trim=0mm 0mm 0mm 0mm, clip]{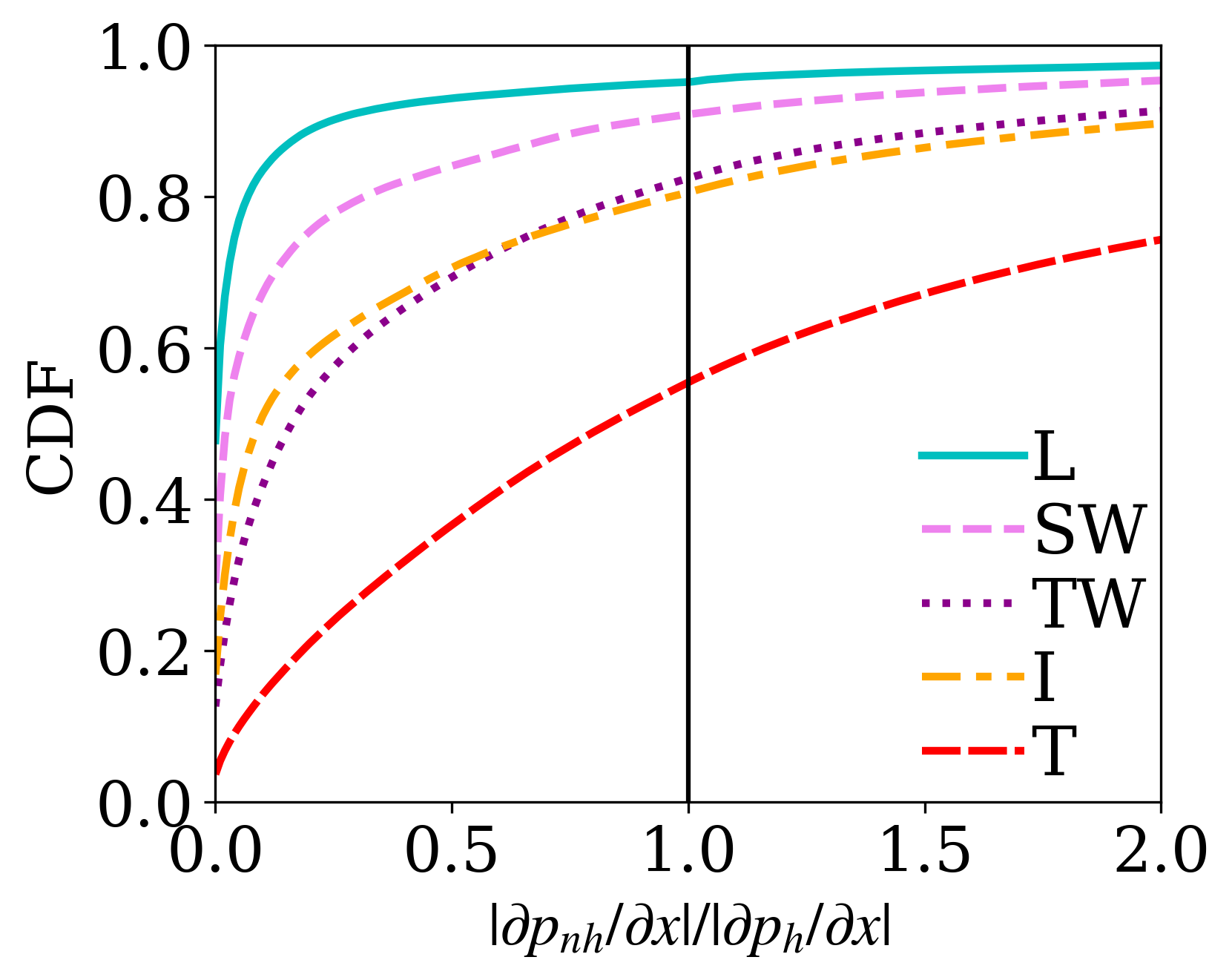}

    \caption{Cumulative Distribution Function (CDF) of the ratio between non-hydrostatic  and  hydrostatic streamwise pressure gradient.}
    \label{fig:dpdx}
\end{figure}

\Cref{fig:dpdx} shows the cumulative density function of the magnitude ratio between $\partial p_h/\partial x$ and $\partial p_{nh}/ \partial x$ in all five datasets L, SW, TW, I and T. We find that the hydrostatic pressure gradient dominates over the non-hydrostatic gradient (i.e.  $|\partial p_{nh}/\partial x|/|\partial p_h/\partial x|<1$  in $\ge 80 \%$ of the data in the L, SW, TW, and I cases, and $\ge 60 \%$ of the data even in the T case.

Non-hydrostatic effects, therefore, play a secondary role in all but the T case, and the two-layer SWE are expected to model adequately the primary two-layer dynamics of SID flows forced by relatively small tilt angles $\theta$.  However, as the flow becomes turbulent (T case), non-hydrostatic effects become important and the applicability of the shallow water model breaks down. Turbulence also creates a third layer of intermediate density (see \cite{ZhuAtoufi2022}, figure 5(e,j)) and the two-layer model also breaks down. For these reasons, we exclude this T case from the analyses of this paper and focus on the L, SW, TW, and I cases.

\section{Derivation of the Taylor-Goldstein equation and solution}
\label{sec:TG2}

In this section, we provide additional details regarding the derivation of the inviscid Boussinesq Taylor-Goldstein equation (TGE)  \eqref{TG_eq}  for a two-layer SID flow and the dispersion relation \eqref{lambda_2L_TG_mt}. 

\subsection{Governing equation}

The linearised inviscid equations for  two-dimensional wave-like perturbations $[\tilde{\psi},\tilde{\rho},\tilde{p}]=[\hat{\psi},\hat{\rho},\hat{p}](z)\exp ik(x-ct)$ around a parallel base flow  $[\mathcal{U},\mathcal{R}](z)$ are
\begin{eqnarray} \label{eq:lin_2dxmom}
\left(\mathcal{U}-c\right) \hat{\psi}'- \hat{\psi} \mathcal{U}'=-\hat{p}-\frac{i}{k} \Ri \sin{\theta} \hat{\rho},
\end{eqnarray}
\begin{eqnarray} \label{eq:lin_2dzmom}
k^2 \left(\mathcal{U}-c\right) \hat{\psi} = -\hat{p}' - \Ri \cos{\theta} \hat{\rho},
\end{eqnarray}
\begin{eqnarray} \label{eq:lin_2denergy}
 \left(\mathcal{U}-c\right) \hat{\rho} - \hat{\psi} \mathcal{R}'=0,
\end{eqnarray}
By taking the $z$ derivative of (\ref{eq:lin_2dxmom}) and using (\ref{eq:lin_2dzmom}) and (\ref{eq:lin_2denergy}) we derive the general TGE as
\begin{eqnarray} \label{eq:TGE-with-sintheta}
&& \left(\mathcal{U}-c\right) \left[\frac{d^2}{d z^2}  -k^2 \right]\hat{\psi} - \mathcal{U}'' \hat{\psi} - \frac{\Ri \cos{\theta} \ \mathcal{R}'}{\mathcal{U}-c} \hat{\psi} = F, \\&& \nonumber
\text{with forcing} \ F=- \frac{i}{k} \Ri \sin{\theta} \left[ \frac{\mathcal{R}'}{\mathcal{U}-c} \hat{\psi}'  + \frac{\mathcal{R}''}{\mathcal{U}-c} \hat{\psi} - \frac{\mathcal{U}' \mathcal{R'}}{\left(\mathcal{U}-c \right)^2} \hat{\psi} \right].
\end{eqnarray}
The streamwise component of the gravitational force $\Ri \sin \theta$ appears multiplied by $i$ such that even if $c$ is real (the waves are stable) increasing the tilt angle will eventually lead to instability. We neglect this effect here. 

For small tilt angles, we assume for simplicity $F=0$. These unforced TG equations under small tilt angles will be the focus of the following stability analysis. The unforced TG equation then (i.e~ $F=0$) is given in  \eqref{TG_eq}.

Taking the base flow as the two-layer piecewise constant profiles in \eqref{eq:baseflow_TGE} leads to $\mathcal{U}'=(u_1-u_2) \ \delta(z)$ and $\mathcal{R}'= (\rho_1-\rho_2)  \ \delta(z)$, where $\delta$ is the Dirac delta function. Since $\mathcal{U}''=\mathcal{R}'=0$ everywhere except at the interface, the TGE becomes trivial
\begin{eqnarray}
   \hat{\psi}''-k^2 \hat{\psi}=0.
\end{eqnarray}

\subsection{Solution with solid top and bottom walls}\label{sec:TGderiv-walls}

In this bounded duct configuration, we take a solution of the form
\begin{eqnarray}
   \hat{\psi}=
    \begin{cases}
        C_1 \sinh{k (h_1-z)} &  0 < z \le h_1, \\
        C_2 \sinh{k (h_2+z)} &  -h_2 \le z < 0,
    \end{cases}
\end{eqnarray}
which satisfies the no-penetration condition at $z=-h_2$ and $z=h_1$ ($\hat{w} =i k\hat{\psi}=0$) modelling the presence of solid walls.  

Following \citet{drazin_reid_2004}, the matching conditions are derived by integrating \eqref{TG_eq} over the neighbouring region of the interface, and by using the integral property of the Dirac delta function, leading to 
\begin{eqnarray} \label{eq:MC_1}
&& \llbracket \frac{\hat{\psi}}{\mathcal{U}-\lambda} \rrbracket_0=0, \\&& 
   \llbracket \mathcal{U}\hat{\psi}'-c \hat{\psi}' \rrbracket_0 + {\Ri  \,  \Delta\rho \, \cos{\theta}} \left(  \frac{\hat{\psi}}{\mathcal{U}-c} \right)_{z=0}=0 , \label{eq:MC_2}
\end{eqnarray}

where we recall that $\Delta\rho\equiv \rho_2-\rho_1$. The first condition (\ref{eq:MC_1}) guarantees continuity of the streamfunction across the interface (and thus the wall-normal velocity). Together, the first and second term in the second condition (\ref{eq:MC_2}) guarantees continuity of the pressure modes $\hat{p}$ based on (\ref{eq:lin_2dxmom}). Note that $\llbracket \mathcal{U}' \hat{\psi} \rrbracket_0 $ vanishes as $\mathcal{U}'=0$ on either side of the interface and $\mathcal{U}' \rightarrow \infty$ at the interface. The last term in (\ref{eq:MC_2}) guarantees continuity of the density modes $\hat{\rho}$ based on (\ref{eq:lin_2denergy}). Using these two conditions we can solve for $C_1$ and $C_2$ leading to the following algebraic system of equations:
\begin{eqnarray}
  && C_1 \sinh{k h_1} (u_2-c) - C_2 \sinh{k h_2} (u_1-c)=0, \\&&
     \left[-\left(u_1-c \right) k \cosh{k h_1} + \Ri \, \frac{\Delta\rho}{2} \, \cos{\theta} \ \frac{\sinh{k h_1}}{u_1-c}\right] C_1 + \\&& \nonumber
     \left[- \left(u_2-c\right) k \cosh{k h_2} + \Ri \, \frac{\Delta\rho}{2}  \, \cos{\theta}  \frac{\sinh{k h_2}}{u_2-c}\right] C_2=0.
\end{eqnarray}
To have a non-trivial solution, the determinant of the above $2\times 2$ system must be zero,  resulting in the dispersion relation \eqref{lambda_2L_TG_mt}.

\subsection{Solution without solid top and bottom walls}\label{sec:TGderiv-nowalls}

In the unbounded configuration, we instead take a solution of the form
\begin{eqnarray}
    \label{eq:solu_unbound_TG}
   \hat{\psi}=C\exp{(-k|z|)},
\end{eqnarray}
satisfying continuous and finite $\hat{\psi}$ for all $z$~\citep{smyth_carpenter_2019}.

Substituting \eqref{eq:solu_unbound_TG} into \eqref{eq:MC_2}, we obtain 
\begin{equation}
    -(u_1-c)^2 k -(u_2-c)^2 k - \Ri \, \Delta\rho\cos\theta=0,
\end{equation}
and thus the dispersion relation for KH waves plotted in figure~\ref{fig:TG_2L_Q}(c)
\begin{equation}\label{eq:TG-disp-rel-no-walls}
   c=\frac{u_1+u_2}{2}\pm \sqrt{\frac{\Ri \,  \frac{\Delta\rho}{2} \cos\theta}{k}-\frac{(u_1-u_2)^2}{4}}.
\end{equation}

The KH instability ($c^I\neq0$) is thus found for 
\begin{equation} 
   k> \frac{2 \, \Ri \,  \Delta\rho \, \cos\theta}{(u_1-u_2)^2}.
\end{equation}

 \bibliographystyle{jfm}
 \bibliography{main}


\end{document}